\documentclass[twocolumn]{aastex631}
\usepackage{wrapfig}
\usepackage{mathptmx}
\usepackage[T1]{fontenc}
\usepackage{ae,aecompl}
\usepackage{graphicx}	
\usepackage{amsmath}	
\usepackage{amssymb}	
\usepackage{subfigmat}
\usepackage{enumerate}
\usepackage{natbib}
\usepackage{verbatim}
\usepackage{tabularx, booktabs}
\usepackage{rotating}
\usepackage{longtable}
\usepackage{ltablex}
\usepackage{threeparttablex}
\usepackage[mathcal]{eucal}

\newcommand{\kms}{km~s$^{-1}$}
\newcommand{\degree}{$^{\circ}$}
\newcommand{\rom}[1]{\uppercase\expandafter{\romannumeral#1}}

\newcommand{\twco}{\mbox{$^{12}$CO}}
\newcommand{\thco}{\mbox{$^{13}$CO}}
\newcommand{\water}{\mbox{H$_2$O}}
\newcommand{\um}{$\mu$m}

\begin{document}

\title{EPISODE II: Variability in the CO and \water\ rovibrational absorption lines in a periodically variable protostar EC 53}


\author[0000-0002-0226-9295]{Seokho Lee}
\affil{Korea Astronomy and Space Science Institute 776 Daedeok-daero, Yuseong-gu, Daejeon 34055, Republic of Korea}
\email{seokholee@kasi.re.kr}

\author[0000-0003-3119-2087]{Jeong-Eun Lee}
\affil{Department of Physics and Astronomy, Seoul National University, 1 Gwanak-ro, Gwanak-gu, Seoul 08826, Republic of Korea}
\affil{SNU Astronomy Research Center, Seoul National University, 1 Gwanak-ro, Gwanak-gu, Seoul 08826, Republic of Korea}
\email{lee.jeongeun@snu.ac.kr}

\author[0000-0002-2523-3762]{Chul-Hwan Kim}
\affil{Department of Physics and Astronomy, Seoul National University, 1 Gwanak-ro, Gwanak-gu, Seoul 08826, Republic of Korea}

\author[0000-0001-8064-2801]{Jaeyeong Kim}
\affil{Korea Astronomy and Space Science Institute 776 Daedeok-daero, Yuseong-gu, Daejeon 34055, Republic of Korea}

\author[0009-0004-7886-9029]{Young-Jun Kim}
\affil{Department of Physics and Astronomy, Seoul National University, 1 Gwanak-ro, Gwanak-gu, Seoul 08826, Republic of Korea}

\author[0000-0002-7154-6065]{Gregory J. Herczeg}
\affil{Kavli Institute for Astronomy and Astrophysics, Peking University, Yiheyuan Lu 5, Haidian Qu, 100871, Beijing, PR China}

\author[0000-0002-6773-459X]{Doug Johnstone}
\affil{NRC Herzberg Astronomy and Astrophysics, 5071 West Saanich Rd, Victoria, BC, V9E 2E7, Canada }
\affil{Department of Physics and Astronomy, University of Victoria, Victoria, BC, V8P 5C2, Canada}

\author[0000-0002-2814-1978]{Giseon Baek}
\affil{Department of Physics and Astronomy, Seoul National University, 1 Gwanak-ro, Gwanak-gu, Seoul 08826, Republic of Korea}
\affil{Research Institute of Basic Sciences, Seoul National University, Seoul 08826, Republic of Korea}

\author[0000-0003-3283-6884]{Yuri Aikawa}
\affil{Department of Astronomy, University of Tokyo, 7-3-1 Hongo, Bunkyo-ku, Tokyo 113-0033, Japan}

\author[0000-0003-1665-5709]{Joel D. Green}
\affil{Space Telescope Science Institute, 3700 San Martin Dr., Baltimore, MD 02138, USA}

\author[0000-0001-8227-2816]{Yao-Lun Yang}
\affil{RIKEN Cluster for Pioneering Research, Wako-shi, Saitama, 351-0198, Japan}
\affil{Department of Astronomy, University of Virginia, Charlottesville, VA 22904, USA}

\author[0000-0001-8822-6327]{Logan Francis}
\affil{Leiden Observatory, Leiden University, PO Box 9513, 2300 RA Leiden, The Netherlands}

\author[0000-0001-7552-1562]{Klaus M. Pontoppidan}
\affil{Jet Propulsion Laboratory, California Institute of Technology, 4800 Oak Grove Drive, Pasadena, CA 91109, USA}

\author[0000-0002-3808-7143]{Ho-Gyu Lee}
\affil{Korea Astronomy and Space Science Institute, 776, Daedeok-daero, Yuseong-gu, Daejeon, 34055, Republic Of Korea}


\author{Takahiro Nagayama}
\affil{Graduate School of Science and Engineering, Kagoshima University, 1-21-35 Korimoto, Kagoshima, Kagoshima 890-0065, Japan}

\author{Yoichi Itoh}
\affil{Center for Astronomy, University of Hyogo 407-2 Nishigaichi, Sayo, Hyogo, 679-5313, Japan}

\author{Tomohito Ohshima}
\affil{Center for Astronomy, University of Hyogo 407-2 Nishigaichi, Sayo, Hyogo, 679-5313, Japan}

\author{Jun Takahashi}
\affil{Center for Astronomy, University of Hyogo 407-2 Nishigaichi, Sayo, Hyogo, 679-5313, Japan}

\author{Jun Toshikawa}
\affil{Center for Astronomy, University of Hyogo 407-2 Nishigaichi, Sayo, Hyogo, 679-5313, Japan}

\author{Tomoki Saito}
\affil{Center for Astronomy, University of Hyogo 407-2 Nishigaichi, Sayo, Hyogo, 679-5313, Japan}

\author{Motohide Tamura}
\affil{Graduate School of Science, The University of Tokyo, 7-3-1 Hongo, Bunkyo-ku, Tokyo 113-0033, Japan}
\affil{Astrobiology Center, 2-21-1 Osawa, Mitaka, Tokyo 181-8588, Japan}
\author{Takayoshi Kusune}
\affil{Department of Physics, Nagoya University, Chikusa-ku, Nagoya 464-8602, Japan}




\begin{abstract}
We present two-epoch JWST NIRSpec and MIRI observations of the young protostar EC~53 (V371~Ser), a periodically variable source with well-characterized quiescent and burst phases. The spectra in both epochs show absorption in the CO overtone ($\sim$2.3\,\um) and fundamental ($\sim$4.6\,\um) bands and the \water\ stretching ($\sim$2.7\,\um) and bending ($\sim$6.0\,\um) modes. We also obtained high-resolution ($R\approx45{,}000$) IGRINS spectra during the burst to constrain the CO overtone line profiles. LTE slab modeling yields gas temperatures of $\sim$1800\,K (CO overtone) and $\sim$1200\,K (CO fundamental), consistent with the overtone tracing hotter gas at smaller radii. The \water\ stretching-mode absorption shows no compelling evidence for variability, and the current JWST CO overtone data do not provide a robust constraint on overtone variability. In contrast, the CO fundamental and \water\ bending-mode features weaken by a factor of $\sim$2 during the burst, which is most naturally explained by continuum changes rather than large variations in absorbing gas. To quantify continuum dilution, we introduce a ``relative veiling'' that treats the quiescent spectrum as an internal reference and measures the change in the continuum excess between the two epochs. 
This formalism yields burst-to-quiescent hot-continuum ratios of $2.9\pm0.2$ for the CO overtone and $1.71\pm0.11$ for the CO fundamental. Using a viscous-disk prescription, these imply representative accretion-rate ratios of $\sim$3.6 and $\sim$2.0, respectively. The differing ratios suggest that inner-disk regions traced at different temperatures, and thus radii, respond differently across the burst cycle, consistent with episodic mass buildup in the inner disk during quiescence followed by more efficient transport through the innermost disk onto the protostar during the burst.

\end{abstract}


\section{Introduction} \label{sec:intro}
Gas instabilities in protoplanetary disks redistribute angular momentum, allowing some material to spiral inward and fuel stellar growth while other gas is transported outward by gaining angular momentum \citep{Turner2014,Kratter2016}.  In addition, magnetically driven disk winds can remove angular momentum from the inner disk, thereby facilitating mass accretion \citep[e.g.,][]{Bai2016}. Mass accretion is the fundamental driver of both stellar growth and circumstellar disk evolution. In the deeply embedded Class~0/I stages, envelope-fed bursts can elevate these rates by orders of magnitude, dominating the luminosity budget and generating transient temperature inversions that reset grain growth and volatile chemistry within the inner few au \citep[e.g.,][]{Hartmann1996, Audard2014, Fischer2022}. Even after envelope dispersal, the monotonic decline of the mass accretion rate ($\dot{M}$) through the Class~II stage regulates disk surface-density profiles and, by extension, the reservoir of solids available for planet formation \citep{Mulders2021}. However, the mass accretion rate is still poorly constrained; $\dot{M}$ may vary with both the time and the disk radius. Furthermore, such an episodically variable accretion rate can affect the physical and chemical conditions of the disk significantly.

Observational evidence for time-variable molecular gas in response to episodic accretion has emerged from studies of eruptive young stellar objects, particularly in more evolved Class~II systems where inner-disk rovibrational molecules are typically observed in emission. 
In EX~Lup, near- and mid-infrared spectroscopy across the 2008 outburst revealed pronounced changes in both the continuum and molecular spectrum: \water\ and OH emission strengthened, while signatures of small organics (e.g., HCN, C$_2$H$_2$, and CO$_2$) disappeared, consistent with enhanced inner-disk heating and UV-driven disk-surface chemistry \citep{Banzatti2012,Smith2025}. 
High-resolution monitoring further showed that the CO fundamental emission evolved with the outburst and included distinct kinematic components whose fluxes decayed together with the infrared continuum \citep{Goto2011}. Post-outburst 3--5~\um\ spectroscopy later indicated an order-of-magnitude decrease in the inner-disk molecular gas mass, comparable to the decline in the stellar accretion rate, supporting a picture in which gas is episodically depleted following enhanced accretion \citep{Banzatti2015}. 
More recent JWST observations further demonstrated that outbursts can leave long-lived chemical signatures, such as enhanced cold \water\ vapor consistent with ice sublimation during snowline recession and delayed freeze-out in disk surface layers \citep{Smith2025}. Time-domain JWST spectroscopy of systems with predictable accretion pulses, such as the eccentric binary DQ~Tau, likewise shows that some hot molecular components correlate with accretion state, while cooler components remain comparatively stable \citep{Kospal2025}. Together, these studies show that time-domain molecular spectroscopy can trace both the physical and chemical response of inner-disk gas to variable accretion. Whether the same diagnostics can be applied in a similar manner to younger, more embedded Class~0/I protostars, however, remains unclear.

Observationally, mass accretion rates have been estimated using several complementary diagnostics. In classical T Tauri stars, UV/optical continuum veiling and hydrogen recombination line fluxes yield $\dot{M}$ via magnetospheric accretion models (see review by \citealt{Manara2023}). In more embedded phases, spectral energy distribution (SED) modeling of IR–sub-mm photometry constrains the total accretion luminosity, from which the mass accretion rate is derived via $L_{\rm acc} \simeq G M_* \dot{M} / R_*$ \citep{Fischer2017}. At high mass accretion rates, such as in FU Ori–type outbursts, this bolometric luminosity is dominated by viscous heating of the inner disk, whereas at more moderate mass accretion rates, magnetospheric shock emission is a dominant contributor \citep{Hartmann2016}. 
While these methods provide useful estimates of the total accretion luminosity, they are subject to significant uncertainties, particularly in embedded phases where envelope reprocessing, source geometry, and variability can affect the observed emission \citep[e.g.,][]{Dunham2014,Offner2012,Fischer2017}. As a result, these approaches  generally provide limited leverage on how the mass accretion rate varies with radius within the inner disk. This limitation motivates the exploration of disk-based diagnostics that can probe accretion-related changes within the inner disk on multiple spatial scales.

CO and \water\ rovibrational lines provide important diagnostics of inner disk conditions that are strongly influenced by accretion processes.
The CO overtone ($\sim$2.3\,\um) and fundamental ($\sim$4.7\,\um) transitions originate from hot gas at different locations within the inner disk atmosphere: the overtone lines trace regions closer to the central protostar and arise from gas at $\sim$2000\,K, while the fundamental lines probe relatively larger radii and arise from gas at $\sim$500--1000\,K. 
In passively heated disks, both transitions are most commonly observed in emission \citep{Najita2003,Brittain2007,Salyk2011,Herczeg2011,Brown2013,vanderPlas2015,Banzatti2022}.
At sufficiently high mass accretion rates, however, enhanced continuum emission from the inner disk can lead these same transitions to appear in absorption, as observed in FU Ori–type systems \citep[see, e.g.,][]{Hartmann1996,Zhu2009,jelee2015,Connelley2018}.
These contrasting behaviors suggest that rovibrational CO and \water\ lines are sensitive to the accretion-powered inner disk environment, motivating their use as probes of accretion-driven variability in eruptive protostars.

EC\,53 (V371~Ser) is a deeply embedded Class~I protostar in the Serpens Main star-forming region \citep[$d$=\,436\,pc;][]{Ortiz-leon2017}, exhibiting quasi-periodic accretion variability with a characteristic timescale of $\sim$1.5\,yr, as revealed by monitoring observations from near-infrared to submillimeter wavelengths \citep{Hodapp1999,Hodapp2012,yhLee2020,Francis2022}. Its bolometric luminosity increases by a factor of 3.3 during the burst phase compared to the quiescent phase \citep{Baek2020}, and ALMA C$^{17}$O 3--2 observations yield a minimum stellar mass of 0.3$\pm$0.1 $M_\odot$ assuming Keplerian rotation \citep{sLee2020}. These characteristics make EC\,53 a particularly well-suited target to investigate accretion-driven variability during the main mass assembly phase through time-resolved infrared spectroscopy.

We find that the CO and \water\ rovibrational lines appear in absorption in our observations toward EC 53 with the James Webb Space Telescope \citep[JWST;][]{Gardner2006}. We also report significant variations in the depths of these lines between the quiescent and burst phases, providing insight into phase-dependent accretion processes within the inner disk and motivate a series of coordinated investigations.
This paper is the second in a series characterizing the accretion-driven variability of EC 53 using JWST spectroscopy. 
Paper~I \citep{jeLee2026} presents a detailed analysis of silicate crystallization in the inner disk. 
Paper~III \citep{Leesj2026} focuses on atomic and molecular emission lines tracing shocks and outflows, such as H$_2$, [Fe~II], and [Ne~II], while Paper~IV \citep{jKim2026} examines ice absorption features associated with the envelope.

Section~\ref{sec:data} describes the observations and data reduction. Section~\ref{sec:results} presents the observed spectral features and their variability. Section~\ref{sec:methods} introduces the analysis framework and Section~\ref{sec:analysis} details the modeling of the gas properties as well as their variability. We discuss the implications of our findings in Section~\ref{sec:discussion}, and summarize our conclusions in Section~\ref{sec:summary}.

 \section{Observations and Data Reduction}\label{sec:data}
\begin{figure*}[htp]
 \centering
 \vspace{-2mm}
 \includegraphics[width=1\textwidth]{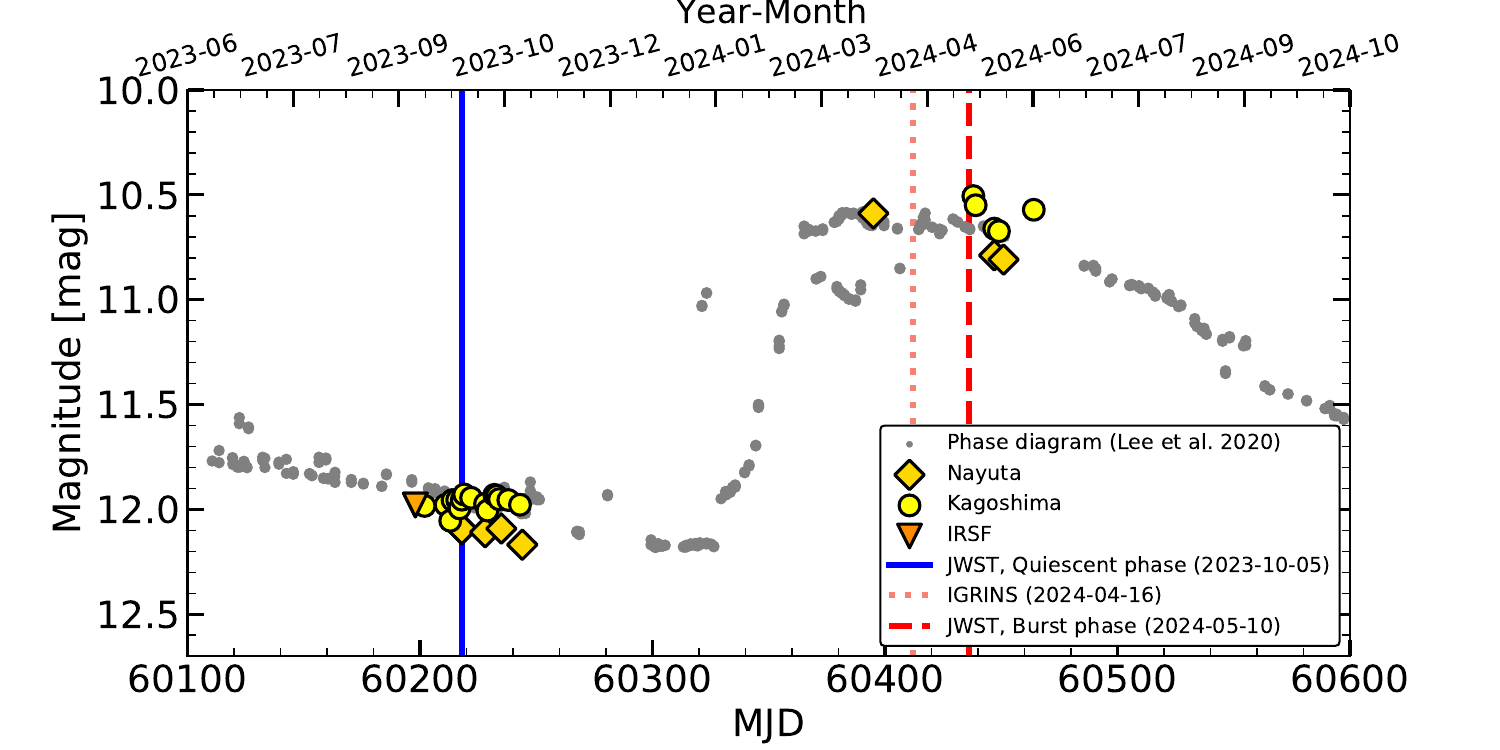}
 \caption{Light curve of EC 53 in Ks band. The gray points indicate the Ks band light curve reconstructed from the phase diagram in \cite{yhLee2020}. The diamond, circle, and triangle symbols represent magnitudes observed with the Nayuta, Kagoshima, and IRSF telescopes, respectively. The vertical solid, dotted, and dashed lines mark the timing of JWST observation during the quiescent phase (2023-10-05), IGRINS observation (2024-04-16), and JWST observation during the burst phase (2024-05-10), respectively. }
\label{fig:lightcurve}
\end{figure*}

As illustrated in Figure~\ref{fig:lightcurve}, our data set combines (i) two integral-field spectra obtained with the JWST—one during the quiescent phase and one during the burst phase—with (ii) two high-resolution ($R\!\approx\!45{,}000$) spectra from the Immersion Grating Infrared Spectrograph \citep[IGRINS;][]{Yuk2010,Park2014}, both acquired during the burst.  To confirm the photometric phase of each spectroscopic visit, we also carried out contemporaneous near-infrared (NIR) $JHK$ monitoring with the Nayuta, Kagoshima, and IRSF telescopes. Together, these observations provide phase-resolved spectroscopy and photometry that anchor our analysis of EC 53’s cyclical variability.

\subsection{JWST}
EC 53 was observed with the Near Infrared Spectrograph \citep[NIRSpec;][]{Boker2022} and the Mid-Infrared Instrument (MIRI) medium-resolution spectrometer \citep[MRS;][]{Wells2015,Labiano2021,Argyriou2023} onboard JWST on October 5, 2023, and May 10, 2024, respectively (PID: 3477, PI: Jeong-Eun Lee).

The NIRSpec IFU observations were conducted using the disperser-filter combinations of G235H-F170LP and G395H-F290LP, providing a resolving power of $\sim$2700 and covering the wavelength range from 1.66 to 5.27\,\um.
For the quiescent-phase observation with NIRSpec IFU, the NRSIRS2RAPID readout mode was used with 7 and 6 groups for G235H-F170LP and G395H-F290LP, respectively, and 2 integrations for each disperser-filter combination. A 4-point dither pattern was applied, resulting in total exposure times of 934 and 817 s for G235H-F170LP and G395H-F290LP, respectively.
For the burst-phase observation, the NRSIRS2RAPID readout mode was also used with 7 groups and 4 integrations for G235H-F170LP and 6 groups and 3 integrations for G395H-F290LP. A 6-point CYCLING dither pattern was adopted, yielding total exposure times of 2801 and 1838 s for G235H-F170LP and G395H-F290LP, respectively.
To improve the pointing accuracy, target-acquisition observations with the F110W filter were performed using the WATA method with 2MASS-J18295360+0117017 as the reference target, employing the NRSRAPID readout mode for both phase observations.

The MIRI MRS observations cover a wavelength range of 4.9--27.9\,\um\ with a resolving power that decreases from $\sim$2783 at 4.9\,\um\ to $\sim$2575 at 27.9\,\um\ \citep{Argyriou2023,Pontoppidan2024,Banzatti2025}. The instrument consists of four channels, each of which is divided into three sub-bands: Short (A), Medium (B), and Long (C). Both visits were obtained with the FASTR1 readout and a four-point dither pattern, but with different group/integration settings: the quiescent-phase visit used 35, 45, and 40 groups in the A, B, and C sub-bands, respectively, with two integrations per band, whereas the burst-phase visit used 18, 16, and 16 groups with five integrations per band. The resulting total exposure times were 2697 s in quiescence and 2908 s in burst. Target acquisition was performed in the FND filter using the FAST readout with 2MASS-J18295360+0117017, and a dedicated background observation was obtained for each epoch.

Both the NIRSpec IFU and MIRI MRS observations were reduced with the JWST Science Calibration Pipeline v1.18.0 \citep{Bushouse2023}, using the jwst\_1364.pmap context of the Calibration Reference Data System \citep{Greenfield2016}.
For the NIRSpec IFU data, the raw exposures were first processed through the Detector1Pipeline, which applied the detector-level corrections and generated the ramp-fitted slope images. The Spec2Pipeline then applied the spectroscopic calibrations, including flat-fielding, wavelength mapping, and flux conversion. Finally, the Spec3Pipeline generated the 3D science data cubes for each filter. 
For the MIRI data, the raw exposures were first processed through the Detector1Pipeline to produce slope images with detector-level corrections, including minor additional corrections for electromagnetic interference noise patterns and reference pixels. The science and dedicated background observations were reduced separately, and the Spec2Pipeline was then used for spectroscopic calibrations, including flat-fielding, wavelength calibration, and flux calibration. Master background subtraction was subsequently performed in the Spec3Pipeline, which produced the final 3D data cubes for each channel and sub-band.

To enable a direct comparison between the two phases, spectra from both instruments were extracted from a common spatial region after astrometric registration.
Astrometric calibration was carried out for the NIRSpec IFU and MIRI MRS cubes of both observing phases. 
For the MIRI data, a spatial offset between the quiescent- and burst-phase observations was corrected using the ALMA submillimeter continuum position as the astrometric reference. Integrated-intensity maps were generated for MIRI Channels 1--4 in both epochs, and their centroid positions were measured using the \textit{Photutils} package \citep{photutils}. The MIRI cubes were then shifted to align these centroids with the ALMA continuum position. For the NIRSpec IFU data, a similar astrometric procedure was applied to the G235H-F170LP and G395H-F290LP datasets, with the alignment based on the peak position of the integrated-intensity maps.

We extracted both epochs using circular apertures centered on EC~53 after applying the astrometric registration described above.
For the NIRSpec IFU cubes, we adopted a fixed aperture diameter of 1.3\arcsec, which is close to the largest practical aperture given the IFU field of view and the small pointing offsets between visits.
For the MIRI MRS cubes, we used wavelength-dependent apertures with diameters of $4\times$ the PSF FWHM following \citet{Law2023}; for the short-wavelength channels in MIRI CH1, where this prescription yields diameters smaller than 1.3\arcsec, we instead fixed the aperture diameter to 1.3\arcsec\ to maintain consistency with the NIRSpec extraction region.

For each MIRI sub-band spectrum, fringe correction was applied to the extracted 1D spectra using the \texttt{fit\_residual\_fringes\_1d} routine provided in the JWST pipeline. While this procedure removes the dominant fringe patterns, low-level residual fringing remains in some wavelength regions and can introduce additional uncertainty, particularly for weak or broad spectral features.
The extracted spectra occasionally contain outlier channels due to bad pixels and residual instrumental artifacts. We therefore masked obvious outliers (isolated channels that deviate strongly from adjacent wavelengths) before constructing the final spectra used for continuum normalization and subsequent analysis.

\subsection{IGRINS}
We observed EC 53 with the 8.1 m Gemini South Telescope using the IGRINS under PID GS-2024A-DD-103 (PI: Jeong-Eun Lee) on 2024 April 16 and 18, during the burst phase, obtaining NIR spectra on two dates to improve the signal-to-noise ratio (SNR).
IGRINS simultaneously provides high-resolution ($R\approx45{,}000$) spectra covering the H and K bands. The observations were carried out using a slit of 0.34\arcsec$\times$ 5\arcsec\ oriented at a position angle of 145\degree, aligned with the outflow direction.
The data were reduced using the IGRINS pipeline \citep{igrins2024}. Telluric lines in EC 53 spectra were removed by dividing the target spectra by an A0V star (HIP 93286) spectrum observed at a similar airmass to EC 53.

\subsection{NIR photometric monitoring}
The luminosity variation of EC~53 is quasi-periodic. To verify that our JWST observations were obtained during the quiescent and burst phases of this cycle, we carried out contemporaneous NIR monitoring observations in the H and $K_s$ bands.
The monitoring data were obtained using the Nishiharima Infrared Camera (NIC) on the 2.0\,m Nayuta Telescope at the Nishi-Harima Astronomical Observatory (Japan), the kSIRIUS camera \citep{Nagayam2024} on the 1.0\,m Kagoshima University Telescope (Japan), and the SIRIUS camera on the 1.4\,m Infrared Survey Facility \citep[IRSF;][]{Nagayama2003} at the South African Astronomical Observatory.
NIR light curves were constructed using aperture photometry, as described in Appendix~\ref{sec:nir_photometry}. Figure~\ref{fig:lightcurve} presents the resulting monitoring data together with the reconstructed $K_s$-band light curve from previous observations \citep{yhLee2020}, demonstrating that the periodic variability of EC~53 persists and confirming that the JWST observations were obtained at the intended phases of the cycle.

\section{Infrared Spectra }\label{sec:results}
\begin{figure*}[ht]
 \centering
\vspace{-2mm}
 \includegraphics[width=1.0\textwidth]{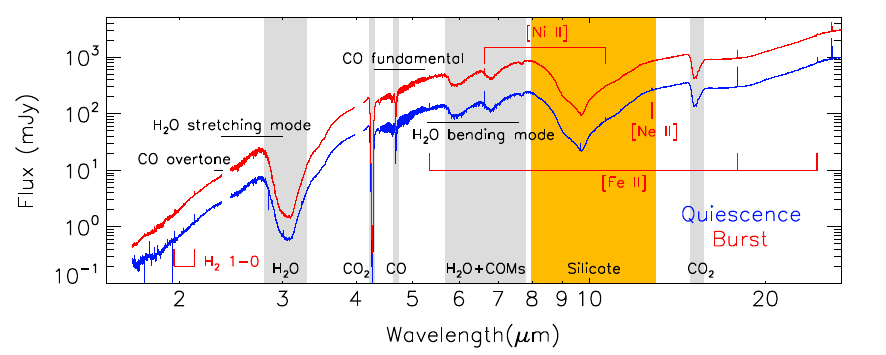}
 \caption{JWST spectra in the burst (red) and quiescent (blue) phases. Among the various spectral features, the CO and \water\ absorption lines are analyzed in this work. The silicate features, emission lines, and ice absorption features are discussed in three separate papers (Papers I, III, and IV). A detector gap at $\sim$2.36-2.47\,\um\ limits the wavelength coverage in both the CO overtone and
\water\ stretching mode regions.}
 \vspace{5mm}
\label{fig:jwst_full}
\end{figure*}

\subsection{Overview of multi-epoch JWST Spectra}\label{subsec:results_overview}
Figure~\ref{fig:jwst_full} presents a comparison of the JWST spectra obtained during the burst and quiescent phases. Both spectra exhibit broad features, including the silicate absorption band near 10\,\micron\ and multiple ice absorption bands. Over most of the covered wavelength range, the continuum level in the burst phase is higher by a factor of about three compared to the quiescent phase. Prominent absorption features are seen in both epochs in the CO fundamental band around 4.6\,\micron\ and in the \water\ bending mode near 6\,\micron, while shallower absorption is present in the CO overtone band around 2.3\,\micron\ and in the \water\ stretching mode around 2.6\,\micron. Several emission lines, including H$_2$ and forbidden transitions of [Fe~II], [Ne~II], and [Ni~II], are detected and analyzed in detail in Paper~III.  No H\,I recombination lines, which are often used as accretion tracers in less embedded systems \citep[e.g.,][]{Tofflemire2025}, are detected within the wavelength range covered by our JWST spectra. At longer wavelengths ($\sim$8--25\,\um), the spectra show low-level fluctuations that may resemble molecular features; however, these structures are not clearly distinguishable in Figure~\ref{fig:jwst_full} and do not provide convincing matches with known molecular transitions over a plausible temperature range. We compared these features with molecular line lists from recent JWST surveys (e.g., the JOYs program; \citealt{vanGelder2024}) and with HITRAN databases \citep{Gordon2026}, but found no robust correspondence. In particular, \water\ rotational transitions expected in this wavelength range are not clearly detected at the current signal-to-noise level. A weak and narrow absorption feature is present near 14.98\,\um\ in both epochs; while its wavelength is broadly consistent with a CO$_2$ transition, the identification remains uncertain at the current signal-to-noise level.

\subsection{Continuum Estimation and Spectral Normalization for the CO and \water\ Bands}
\label{subsec:prep_spectra}
\begin{figure*}[ht!]
 \centering
 \vspace{+2mm}
 \includegraphics[width=0.48\textwidth]{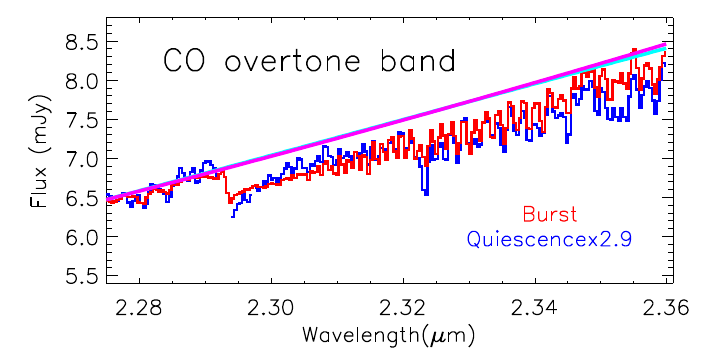}
 \includegraphics[width=0.48\textwidth]{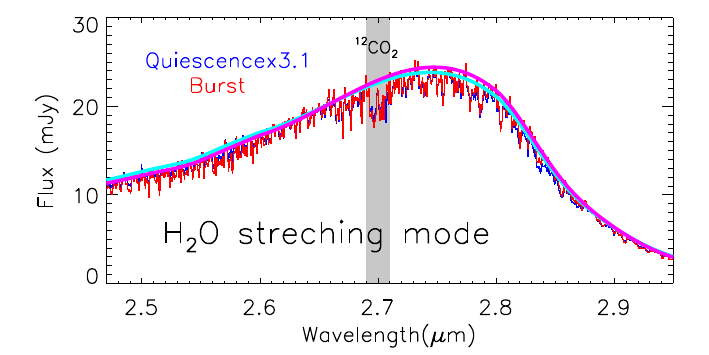}
 \includegraphics[width=0.48\textwidth]{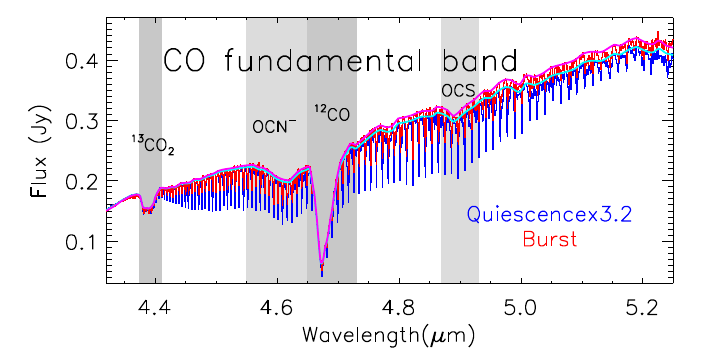}
 \includegraphics[width=0.48\textwidth]{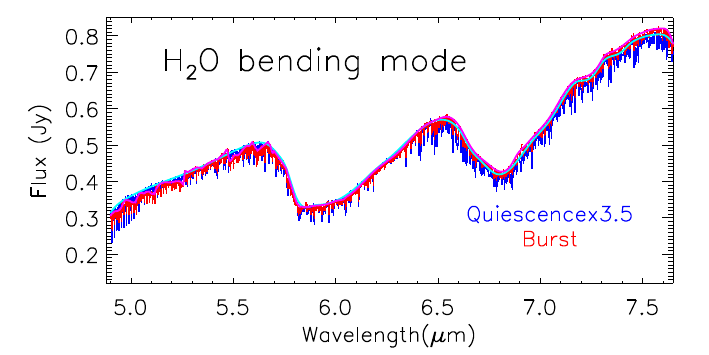}
 \vspace{+5mm}
 \caption{Zoom-in views of the CO and \water\ rovibrational absorption bands toward EC~53, extracted from the full JWST spectra shown in Figure~\ref{fig:jwst_full}. The red and blue curves denote the burst and quiescent phases, respectively. To align the continuum levels and better highlight variations in the molecular absorption features,  the quiescent-phase spectra are scaled by factors of 2.9 and 3.1 in the CO overtone and \water\ stretching mode panels, and by factors of 3.2 and 3.5 in the CO fundamental and \water\ bending mode panels, respectively. In all panels, magenta and cyan curves indicate the adopted continuum levels for the burst and quiescent phases, obtained by interpolating channels that are free of, or only weakly affected by absorption lines. 
In the top-right panel, the \water\ stretching-mode absorption is superposed on broad \water\ ice absorption near $\sim$3\,\um\ (as also indicated in Figure~\ref{fig:jwst_full}) and is additionally contaminated by a weak, locally blended $^{12}$CO$_2$ ice feature (gray shaded region). In the CO fundamental band (bottom-left), the gray shaded regions mark prominent \textit{local} ice absorption features, including $^{13}$CO$_2$ ($\sim$4.39\,\um), OCN$^{-}$ ($\sim$4.6\,\um), \twco\ ($\sim$4.675\,\um), and OCS ($\sim$4.90\,\um). In the \water\ bending-mode panel (bottom-right), the spectrum is superposed on broad ice absorption near 6\,\um\ and 7\,\um\ indicated in Figure~\ref{fig:jwst_full}. }
\label{fig:jwst_zoom}
\end{figure*}

Figure~\ref{fig:jwst_zoom} displays zoomed-in views of the CO overtone and fundamental bands and the \water\ stretching and bending modes extracted from the full spectra in Figure~\ref{fig:jwst_full}. To align the continuum levels and better highlight variations in the molecular absorption features, the quiescent-phase spectra are scaled by factors of 2.9 and 3.1 in the CO overtone and \water\ stretching mode panels, and by factors of 3.2 and 3.5 in the CO fundamental and \water\ bending mode panels, respectively. The gray shaded regions mark wavelength intervals strongly affected by ice absorption. 
To enable a uniform comparison of absorption depths between epochs, we estimate the local continuum separately for the burst and quiescent phases in each band and normalize the spectra by these continua. Using the band-dependent procedure described below, we determine continuum levels for the burst and quiescent phases; these are shown by the magenta and cyan curves in each panel. Unless otherwise noted, all normalized spectra in Figures~\ref{fig:co_overtone_jwst}–\ref{fig:h2o_bending_obs_model} are obtained by dividing the observed flux by these continuum estimates. 

Reliable measurement of the rovibrational absorption features requires an accurate estimate of the underlying continuum in each spectral region. Because the spectra contain a mixture of molecular absorption lines and, in several bands, broad solid-state features, the continuum cannot be determined through simple polynomial fitting alone. Instead, we adopt a band-dependent, model-assisted procedure to identify continuum anchor points and construct smooth baselines.

We first describe the continuum determination for the three bands strongly affected by ice absorption as well as rovibrational lines: the \water\ stretching mode, the CO fundamental band, and the \water\ bending mode. In these regions, broad ice absorption features from CO$_2$, \twco, \water, and other species overlap with the rovibrational lines, making it difficult to identify line-free channels directly from the observed spectra. To locate wavelengths in which the expected gas opacity is negligible, we generate local thermal equilibrium (LTE) slab models for the relevant molecular species and use them to select initial continuum anchor points, where the modeled absorption depth is smaller than the uncertainty level given in Table~\ref{tb:snr}. These points are therefore considered minimally affected by absorption and are used as continuum anchor points. However, using these automatically selected points alone leaves residual structure from the broad ice features in the derived continuum. To better suppress the influence of the ice absorption, we manually add or remove anchor points to refine the continuum placement. The continuum is then constructed using a spline interpolation through these selected points, which provides a smooth representation of the underlying continuum without introducing artificial features. As discussed in Section~\ref{subsec:analysis_lte}, the ice features (gray shaded region in Figure~\ref{fig:jwst_zoom}) are not included in the LTE fitting and thus introduce an additional source of systematic uncertainty, leaving the continuum near these wavelengths intrinsically less secure.

The CO overtone region near 2.3\,\micron\ is treated separately because its wavelength coverage is much narrower than that of the other bands. In this case, the continuum is constrained using two surrounding intervals that are free of both molecular lines and ice features: (1) the clean region immediately shortward of the 2–0 overtone bandhead, and (2) a portion of the \water\ stretching mode region around 2.7\,\micron. The continuum level in the \water\ stretching mode band is first determined using the procedure described above, and this baseline is then adopted as an external constraint when fitting a second-order polynomial across the CO overtone band. 

To quantify the noise properties near each absorption band, we use the nominal line-free intervals listed in Table~\ref{tb:snr}. Although the JWST pipeline provides pixel-level uncertainty estimates, these are not directly applicable to our spectra, which are extracted using a fixed aperture at each wavelength. We therefore estimate the noise empirically from the extracted spectra to ensure consistency with our measurement procedure. A 20-channel median filter is applied to estimate the local continuum, and the standard deviation of the residuals defines the continuum sensitivity. The corresponding uncertainty in the normalized spectra is obtained by dividing this sensitivity by the continuum level. Because the nominal line-free intervals still contain weak molecular contributions, the resulting sensitivities are conservative and may overestimate the true noise level. 

\begin{deluxetable*}{cccccc}
\tablecaption{Sensitivity and RMS \label{tb:snr}}
\tablehead{
\colhead{\bf Transition} & \colhead{\bf Line-Free channels$^a$ (\um)} & \multicolumn{2}{c}{\bf Sensitivity$^b$ (mJy) } &\multicolumn{2}{c}{\bf Uncertainty$^c$ }  \\
\colhead{} & \colhead{} & \colhead{\bf Quiescence}& \colhead{\bf Burst}& \colhead{\bf Quiescence}& \colhead{\bf Burst} }
\startdata
 CO overtone band     & [2.17,2.22] & 0.020 & 0.013 & 0.012  & 0.0027 \\
 \water\ stretching mode& [3.3,3.4] & 0.049 & 0.15 & 0.0065  & 0.0070  \\
 CO fundamental band& [4.28,4.345]  & 0.081 & 0.51 & 0.0018  & 0.0037  \\
 \water\ bending mode& [6.20-6.33]  & 1.48  & 4.15 & 0.011   & 0.0091 
 \enddata
\tablecomments{$^a$ Line-free wavelength intervals used for sensitivity estimation. \\
       $^b$1$\sigma$ uncertainty.\\
        $^c$1$\sigma$ uncertainty level in the normalized flux spectrum.}
\end{deluxetable*}

\subsection{JWST normalized spectra}
\begin{figure*}[htp]
 \centering
 \vspace{-2mm}
 \includegraphics[width=1.0\textwidth]{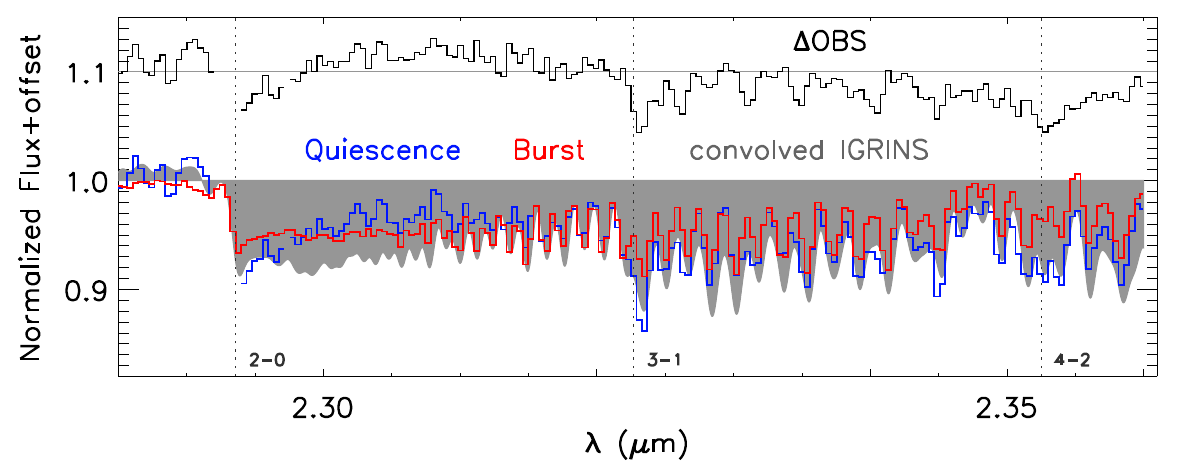}
  \vspace{-5mm}
 \caption{CO overtone absorption features toward EC~53 during the quiescent (blue) and burst (red) phases observed with JWST/NIRSpec. The black curve above the spectra shows the difference between the burst and quiescent spectra. The gray filled spectrum represents the IGRINS burst-phase spectrum convolved to the NIRSpec spectral resolution for comparison. The dotted vertical lines mark the wavelengths of the CO overtone bandheads. Data points near the 2--0 bandhead in the quiescent-phase spectrum were removed due to bad pixels identified during data reduction.}
\label{fig:co_overtone_jwst}
\end{figure*}

Figure~\ref{fig:co_overtone_jwst} shows the normalized CO overtone absorption during the two JWST epochs. The gray filled spectrum is the IGRINS burst-phase spectrum convolved to the NIRSpec resolution for comparison. In this comparison, only the spectral resolution has been matched; differences in seeing, slit width, and aperture between the IGRINS and JWST observations are not explicitly accounted for. Even allowing for a simple scaling factor between the NIRSpec aperture and the IGRINS slit, the convolved IGRINS profile does not reproduce the JWST spectra. The origin of this discrepancy therefore remains uncertain and may reflect a combination of  uncertainties in the continuum estimation of the JWST spectra, unaccounted aperture/seeing effects, residual fringing, subtle differences in instrumental response, and other calibration uncertainties. The quiescent $v=2$--0 bandhead region is affected by bad pixels within the JWST extraction aperture and is therefore masked in the final spectrum.

Between the two JWST epochs, the CO overtone absorption profiles are broadly similar, with  noticeable but localized residual differences in limited wavelength intervals at the $\sim$4--5$\sigma$ level. To assess whether these differences could be explained by temperature variations, we examine the relative behavior of the bandheads and long-wavelength tails. Temperature changes are expected to produce systematic variations: higher temperatures enhance the bandheads, while lower temperatures enhance the bandtails. However, the observed residuals do not follow such a coherent pattern.   
The apparent enhancement of the quiescent $v=2$--0 bandhead is affected by bad pixels and is not reliable. No corresponding enhancement is seen at the $v=3$--1 bandhead. Instead, some residual differences appear at slightly longer wavelengths, offset from the bandhead positions. 
These inconsistencies indicate that the observed variations cannot be explained by a simple temperature change. It remains unclear whether they reflect intrinsic variability or residual calibration uncertainties.
We therefore treat the CO overtone variability as inconclusive in the present data set.

\begin{figure*}[ht!]
\centering
 \vspace{-2mm}
 \includegraphics[width=1.0\textwidth]{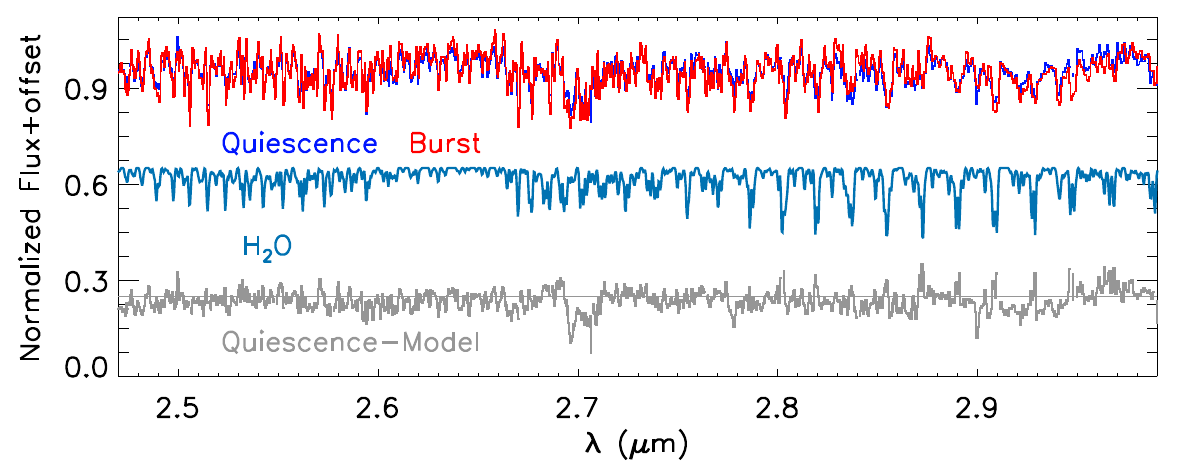}
 \vspace{-5mm}
 \caption{\water\ stretching mode absorption features during the quiescent (blue) and burst (red) phases observed with JWST/NIRSpec. The synthesized spectrum for the quiescent phase is shown below the observations and is generated using only the \water\ component. The model parameters used for the synthesis are listed in Table~\ref{tb:jwst_model}. The residual between the observed and synthesized quiescent spectra is displayed at the bottom in gray.}
\label{fig:h2o_2.7um}
\end{figure*}

The \water\ stretching mode absorption around 2.7\,\micron\ is shown in Figure~\ref{fig:h2o_2.7um}. In this wavelength range the rovibrational lines are blended with a broad $^{12}$CO$_2$ ice feature near 2.7\,\micron, which is difficult to remove clearly, so we retain the ice absorption in the normalized spectra. The normalized burst and quiescent spectra almost perfectly overlap, indicating that the depth of the \water\ stretching mode absorption remains nearly unchanged between the two epochs within the measurement uncertainties.

\begin{figure*}[htp]
 \centering
 \vspace{-2mm}
 \includegraphics[width=1.0\textwidth]{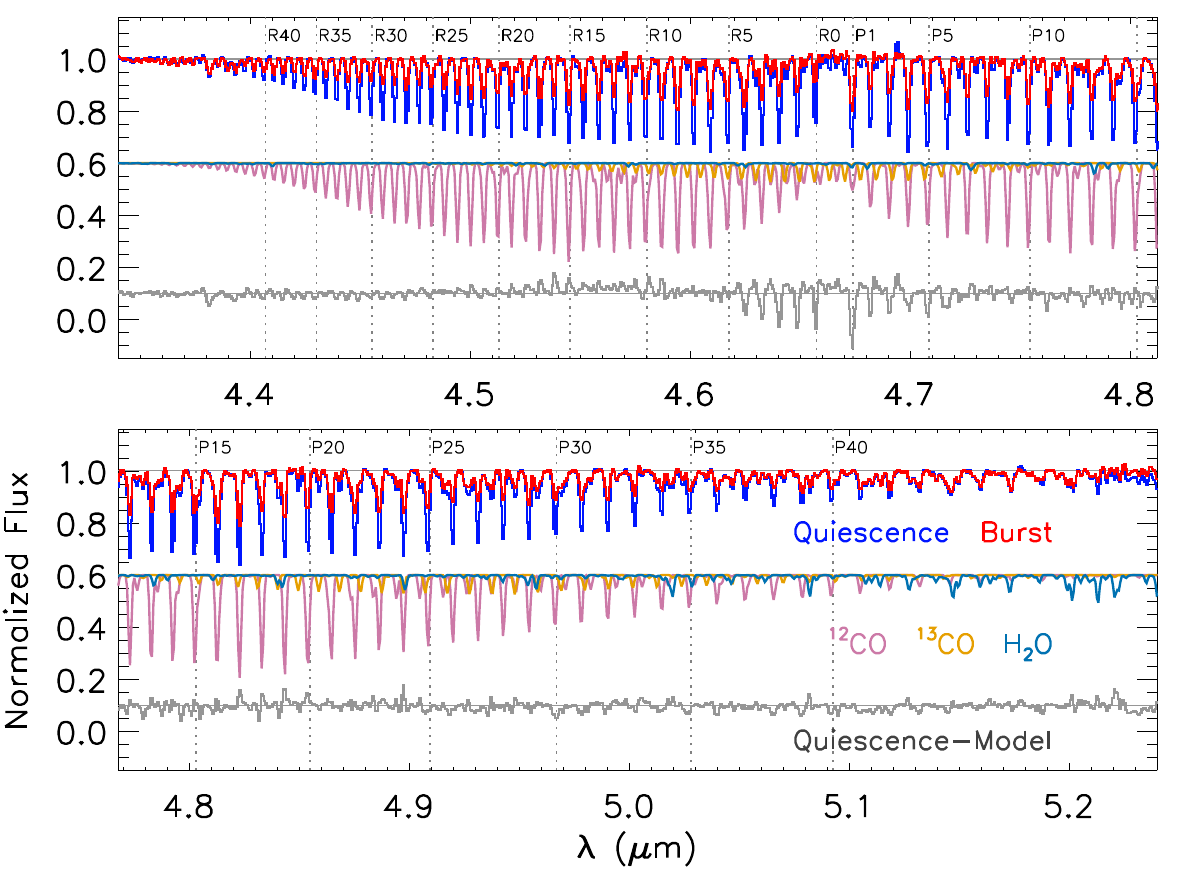}
 \vspace{-5mm}
 \caption{Absorption features in the CO fundamental band during the quiescent (blue) and burst (red) phases observed with JWST/NIRSpec. The synthesized spectrum of the quiescent phase is shown below the observations, where the individual contributions from \twco\ (pink), \thco\ (orange), and \water\ (pastel blue) are overplotted. The model parameters used for the synthesis are summarized in Table~\ref{tb:jwst_model}.  
The residual between the observed and synthesized spectra in the quiescent phase is shown at the bottom in gray. Vertical dotted lines mark the wavelengths of the individual CO $v$=1-0 rotational transitions. }
\label{fig:co_fudamental_obs_model}
\end{figure*}
\begin{figure*}[htp]
 \centering
 \vspace{-2mm}
 \includegraphics[width=1.0\textwidth]{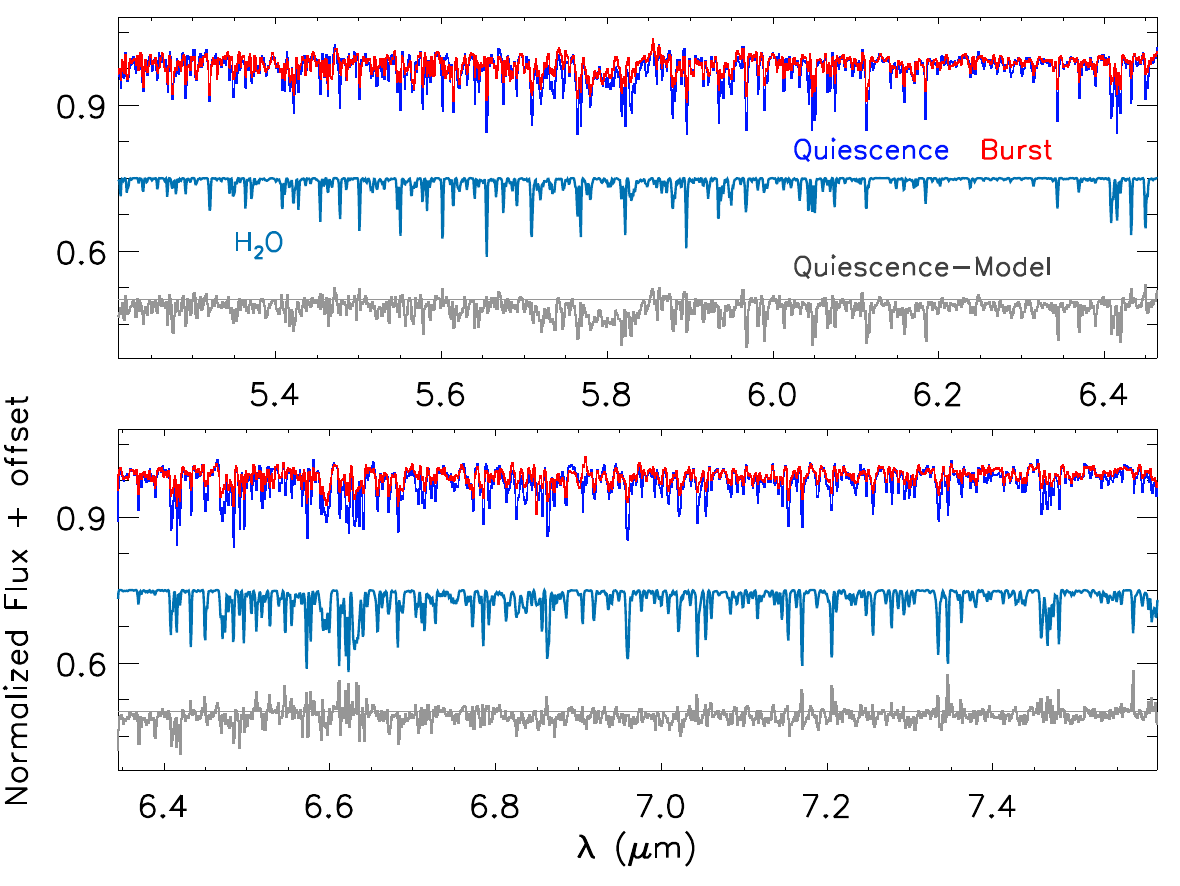}
 \vspace{-5mm}
 \caption{Same as Figure~\ref{fig:co_fudamental_obs_model}, but for the \water\ bending mode absorption features. The quiescent spectrum is modeled using only the \water\ component.}
\label{fig:h2o_bending_obs_model}
\end{figure*}

Figures~\ref{fig:co_fudamental_obs_model} and \ref{fig:h2o_bending_obs_model} present the normalized CO fundamental band and \water\ bending mode absorption, respectively.  
Although the MIRI spectrum overlaps with the NIRSpec coverage down to $\sim$4.9\,\um, the two spectra are consistent in this region after normalization. We therefore show only wavelengths longward of $\sim$5.25\,\um\ in Figure~\ref{fig:h2o_bending_obs_model}, where the molecular features are more clearly separated and the \water\ bending-mode absorption becomes prominent.  
In the CO fundamental region, the strongest features are the \twco\ $v=1$–0 lines. Shallower absorptions with normalized flux levels around 0.9 arise from a combination of \twco\ $v=2$–1, \thco\ $v=1$–0, and \water\ bending mode transitions. 

A clear difference is seen between the two epochs. Compared to the quiescent phase, the burst phase CO lines are weaker by roughly a factor of two in depth, even though the continuum in this band brightens by more than a factor of three. A similar trend is seen in the \water\ bending mode: the strongest \water\ lines in this wavelength region are significantly shallower in the burst spectrum than in the quiescent spectrum, again by approximately a factor of two. These qualitative behaviors provide the observational basis for the quantitative modeling presented in Section~\ref{sec:analysis}. 

\subsection{IGRINS Spectra} \label{subsec:resultsIgrins}
\begin{figure*}[htp]
 \centering
 \vspace{-2mm}
 \includegraphics[width=1.0\textwidth]{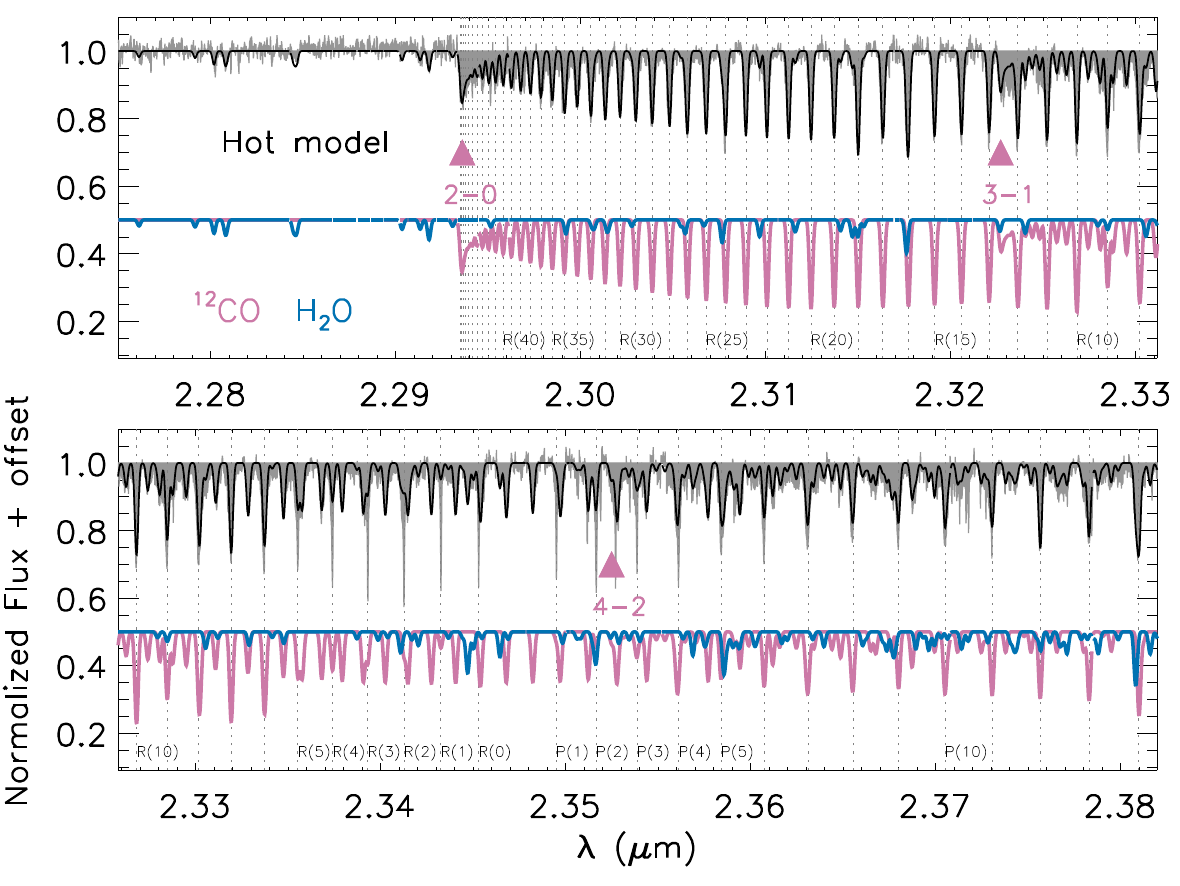}
  \vspace{-5mm}
 \caption{Absorption features in the observed IGRINS spectrum (filled gray) and the synthesized model (color) with offset: hot  \twco\ (pink), hot \water\ (blue), cold \twco\ (orange). The combined contribution of hot \twco\ and \water\ is over-plotted with the black spectrum on the observed spectrum. The model parameters are listed in Table~\ref{tb:co_overtone_model}. The vertical dotted lines indicate the wavelengths of \twco\ $\nu$=2--0 transitions. Pink upward triangles mark the CO bandheads at 2--0 (2.294\,\um), 3--1 (2.323\,\um), and 4--2 (2.353\,\um). Excluding the 2--0 lines at wavelengths longer than the 3--1 and 4--2 bandheads, absorption features with depths greater than 0.9 and less than 0.9 correspond to the 3--1 and 4--2 lines, respectively. \thco\ lines are not detected.}
\label{fig:co_overtone_igrins}
\end{figure*}

Figure~\ref{fig:co_overtone_igrins} presents the CO overtone absorption profiles obtained with IGRINS. Overtone transitions of \twco\ from $v=2$--0 to $v=4$--2 are detected, along with rovibrational absorption lines of the \water\ stretching mode. The strongest absorption features arise from the $v=2$--0 band. Although the spectral range includes \thco\ overtone transitions, they are not detected due to the limited SNR.

\begin{figure}[htp]
 \centering
\includegraphics[width=0.48\textwidth]{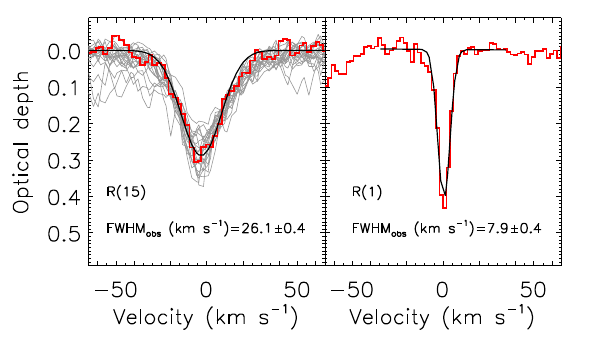}
  \vspace{-5mm}
 \caption{Optical-depth profiles of CO overtone absorption observed with IGRINS. Left: broad component traced by individual $v=2$--0 transitions from R(10) to R(30), shown in light gray. The R(15) transition, which exhibits the narrowest profile, is highlighted in red, with a Gaussian fit overplotted (black), yielding $FWHM_{\rm obs}=26.1\pm0.4$\,\kms. 
Right: narrow component measured from the R(1) transition, with a Gaussian fit yielding $FWHM_{\rm obs}=7.9\pm0.4$\,\kms. The narrow component is unresolved at the IGRINS spectral resolution ($\sim$7.5\,\kms).  }
\label{fig:igrins_fwhm}
\end{figure}

The CO overtone absorption consists of two components: a broad component and a narrow component. 
The line width of the broad component is determined from individual $v=$\,2--0 transitions between R(10) and R(30) as shown in Figure~\ref{fig:igrins_fwhm}. The observed widths span $FWHM_{\rm obs}\sim$26--33\,\kms. Among these, the narrowest profile is found for the R(15) transition, with $FWHM_{\rm obs}=26.1\pm0.4$\,\kms.  
Assuming the nominal IGRINS spectral resolution ($\sim$7.5\,\kms; \citealt{Yuk2010,Park2014}), a simple deconvolution yields an intrinsic width of $\sim$25\,\kms\ for the broad component. However, the effective spectral resolution can vary depending on slit illumination, target centering, and position within the spectral order, so this value should be regarded as an estimate.
The narrow component is measured from the R(1) line as shown in Figure~\ref{fig:igrins_fwhm}. It is unresolved at the spectral resolution of IGRINS ($\sim$7.5\,\kms), and therefore its intrinsic line width cannot be directly measured. We report the observed width ($FWHM_{\rm obs}\sim7.9$\,\kms) as an upper limit.
The measured line-center velocities of the broad and narrow components are offset from the systemic velocity by $-3.0\pm0.2$~\kms\ and $+0.6\pm0.2$~\kms, respectively, indicating that both components are consistent with originating near the systemic velocity within the uncertainties.

\section{Analysis Methods}\label{sec:methods}
In this section, we develop a general analysis framework that connects the multi-epoch CO and \water\ rovibrational spectra to the thermal structure of the inner disk and, ultimately, to the burst-to-quiescent mass-accretion-rate ratio. Section~\ref{subsec:method_twolayer} introduces a two-layer disk geometry in which the absorption lines form in a disk atmosphere that is backlit by a hot midplane continuum emission and diluted by additional emission from a warm midplane; this geometry provides the basic radiative-transfer context needed to interpret the observed absorption profiles. In Section~\ref{subsec:method_lte}, we use simple LTE slab models to characterize the line-forming gas at each epoch, deriving gas temperatures and column densities that describe the absorbing gas independently of the observed variability between phases; these gas temperatures also provide a lower bound on the temperature of the hot midplane continuum source. Building on the two-layer geometry, Section~\ref{subsec:method_veiling_ratio} introduces the relative veiling $\mathcal{V}$, a generalization of classical veiling that explicitly separates the effects of continuum brightening from changes in gas column density when comparing absorption-line profiles between epochs. Finally, Section~\ref{subsec:method_hot} interprets the hot midplane continuum emission in terms of a viscously heated inner disk, supplemented where necessary by magnetospheric accretion shocks, which allows us to convert the inferred continuum emission variations into constraints on the mass accretion rate.

\subsection{Two-layer disk model}\label{subsec:method_twolayer}
\begin{figure*}[htp]
 \centering
 \includegraphics[width=1\textwidth]{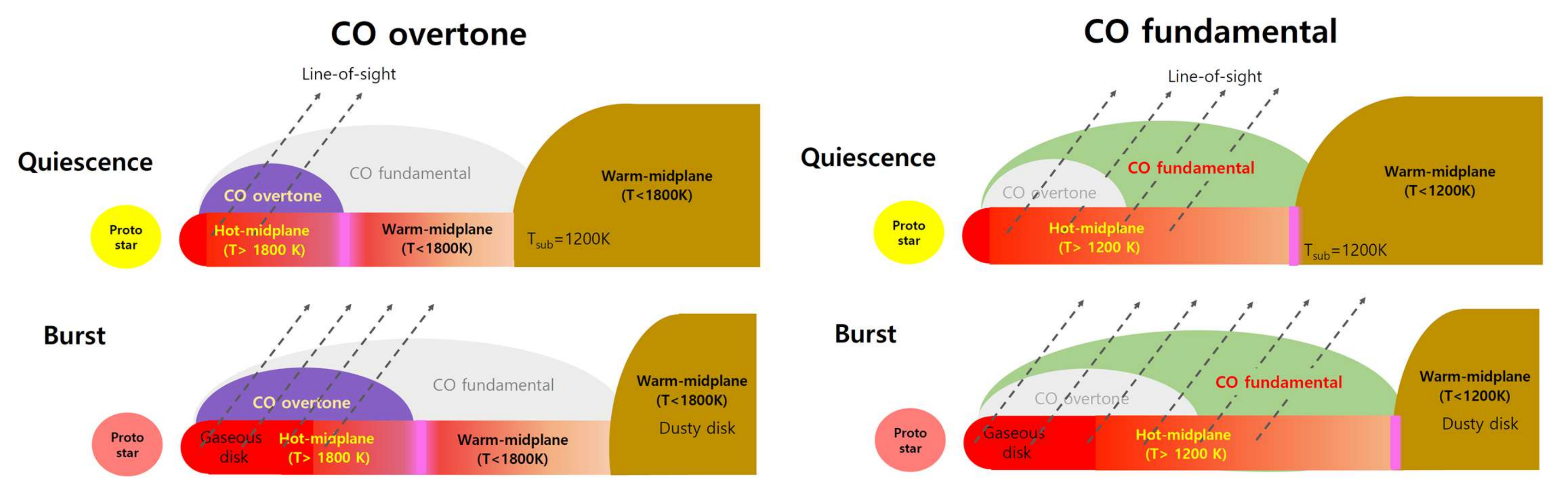}
 \caption{Schematic illustration of the CO absorption-line geometry during the quiescent (top) and burst (bottom) phases for the CO overtone (left) and CO fundamental (right) transitions.  
The dashed arrows indicate the line of sight from the hot midplane continuum source, through the CO absorbing gas, to the observer. 
In each panel, the pink vertical bar marks the boundary between the hot and warm midplane continuum components, defined relative to the characteristic temperature of the absorbing gas. 
For the overtone, this boundary corresponds to a temperature of $\sim$1800\,K, while for the fundamental it corresponds to $\sim$1200\,K, as derived from the LTE slab analysis (Section~\ref{subsec:analysis_lte}). 
During the burst phase, the spatial extent of the hot midplane increases in both cases.}
\label{fig:two_layer}
\end{figure*}

The observed absorption profile can be explained with the two-layer disk model shown in Figure~\ref{fig:two_layer}. The continuum emission arises from two thermally distinct components: (i) a viscously heated gaseous (and partially dusty) midplane and (ii) a passively heated dusty disk. CO overtone and fundamental absorptions form in the disk atmosphere above the disk midplane with these two thermal components.  In this paper, we use the term ``midplane'' in a simplified, phenomenological sense to represent the dominant continuum-emitting components that backlight and/or dilute the molecular absorption, rather than implying a unique physical location within the disk.

For each transition, we divide the midplane continuum source into two components according to the characteristic temperature of the absorbing gas, as illustrated in Figure~\ref{fig:two_layer}.  The first is the hot midplane, whose temperature exceeds that of the absorbing gas and therefore supplies the background continuum emission for the absorption lines. The second is the warm midplane, which is either cooler than the line-forming gas or lies outside the line of sight to the absorbing gas, and thus contributes additional infrared continuum that dilutes the apparent absorption without directly participating in the line formation.  

As shown schematically in Figure~\ref{fig:two_layer}, the boundary between these two components is not a fixed structural layer, but is defined relative to the characteristic temperature of the absorbing gas. This division differs between the CO overtone and CO fundamental transitions because they probe gas at different temperatures (see Section~\ref{subsec:analysis_lte}). For the overtone, the hot midplane is confined to the innermost gaseous disk, while for the fundamental it extends to larger radii, near the dust sublimation regime. During the burst phase, the spatial extent of the hot midplane increases in both cases, reflecting the enhanced continuum emission.

Vertically, we assume the disk is self-similar \citep{Liu2022}: once the local midplane temperature is specified, the emergent continuum level, $F_{\lambda}^{\mathrm{hot}}$, and the flux at the line, $F_{\lambda}$, are uniquely determined. 
In such a geometry, changes in either the properties of the absorbing gas or the emitting area and temperature of the midplane can modify the observed absorption depths; which of these effects dominates for EC~53 is assessed in Sections~\ref{sec:analysis} and \ref{sec:discussion}.

\subsection{LTE Analysis} \label{subsec:method_lte}
The LTE slab analysis provides the basic physical parameters of the absorbing gas, including its gas temperature and column density. These quantities allow us to quantify any differences between the two epochs. Because the absorbing layer must lie in front of the hot midplane, the gas temperature also sets a natural lower bound on the temperature of the underlying hot midplane. In this way, the LTE-derived properties characterize both the intrinsic state of the line-forming gas and the thermal constraints on the midplane continuum source, independent of whether the observed absorption depths differ between the two phases.

To obtain these gas properties, we synthesize the observed absorption bands using a simple LTE slab model. In this approach, we assume a single-temperature absorbing layer characterized by a Gaussian intrinsic line profile. The level populations are computed under LTE using the Boltzmann distribution for a given temperature and column density, and the total optical depth is obtained by summing over all transitions, including line overlap within each molecule. The absorption spectrum is then constructed as $\exp(-\tau)$ and convolved with the instrumental spectral resolution. A detailed description of the model is provided in Appendix~\ref{app:lte_model}.

The spectroscopic data are taken from the HITRAN database \citep{Gordon2026}. For simplicity, we fix the intrinsic line width to 25\,\kms, based on the line widths measured from the high-resolution spectra discussed in Section~\ref{subsec:resultsIgrins}. The resulting spectra are then convolved with the spectral resolutions of IGRINS \citep{Yuk2010,Park2014}, NIRSpec \citep{Boker2022}, and MIRI \citep{Labiano2021}. Best-fitting parameters are obtained via $\chi^{2}$ minimization, using only spectral channels that are not strongly affected by the broad ice absorption features. Note that we use the $\chi^2$ minimization primarily as a relative fitting metric rather than as a strict goodness-of-fit statistic.

\subsection{Relative veiling\,($\mathcal{V}$)}\label{subsec:method_veiling_ratio}

Our spectra sample EC 53 at two distinct epochs with substantially different continuum levels (see Figures~\ref{fig:jwst_full} and \ref{fig:jwst_zoom}). In such a system, changes in the apparent depth of absorption lines can arise either from intrinsic variations in the absorbing gas or from additional continuum emission that dilutes the lines. To interpret the line-depth variability in a physically meaningful way, we therefore require a formalism that explicitly separates the effects of continuum brightening from changes in gas column density.  In practice, a direct estimate of the absolute infrared veiling at each epoch is not feasible for EC~53 because it would require an intrinsic (unveiled) template spectrum at the same wavelengths and resolution, which is not available for this embedded protostar. We therefore focus on a quantity that can be constrained internally from the two-epoch data.

Classical infrared veiling, $r_\lambda$, is widely used to quantify excess continuum emission in T~Tauri stars by comparing spectra to a non-accreting photospheric template. In that framework, $r_\lambda$ is defined for a single epoch and is anchored to an external reference spectrum. In this work we generalize this concept to time-variable disk spectra by introducing a ``relative veiling'', $\mathcal{V}_\lambda$, that measures the change in continuum dilution between two epochs directly from the normalized absorption profiles. This approach is motivated by the empirical observation that the two-epoch spectra are related in a manner consistent with the classical veiling relation in normalized-flux space, as expected if an additional continuum component dilutes the absorption lines. Unlike classical veiling estimates that rely on an external unveiled template spectrum, $\mathcal{V}_\lambda$ can be inferred directly from the transformation between the two-epoch normalized spectra and remains well-defined even if modest veiling is present in both epochs, enabling a consistent epoch-to-epoch quantification of continuum dilution in the absence of a suitable external template. A similar concept of relative veiling has recently been used to quantify epoch-to-epoch continuum variations in time-variable accretion spectra (e.g., \citealt{Herczeg2023}). In that work, relative veiling is primarily employed as an empirical measure of spectral variability. Here, we instead examine the mathematical structure of this quantity and use it to derive a physically motivated decomposition of the continuum emission, as developed below.

At a given epoch $i$ (with $i = Q$ for the quiescent phase and $i = B$ for the burst phase), the observed absorption profile is diluted by continuum from a warm midplane. When this additional continuum emission,  $F_\lambda^{\mathrm{warm}}(i)$, is present alongside the hot midplane continuum emission associated with the absorption, $F_\lambda^{\mathrm{hot}}(i)$, the observed absorption profile, $\tilde{F}_{\lambda}(i)$, appears shallower than the intrinsic absorption profile, $\tilde{F}_{\lambda,o}(i)=F_{\lambda}(i)/F_\lambda^{\mathrm{hot}} (i)$:
\begin{eqnarray}\label{eq:veiling}
       \tilde{F}_{\lambda}(i) &=& \frac{F_{\lambda}(i) + F_\lambda^{\mathrm{warm}}(i)}{F_\lambda^{\mathrm{hot}} (i) + F_\lambda^{\mathrm{warm}}(i)}\\ \nonumber
    &=&  \frac{\tilde{F}_{\lambda,o}(i)+ r_{\lambda}(i)}{1 + r_{\lambda}(i)},
\end{eqnarray}
where $F_\lambda(i)$ is the flux at the absorption line at epoch $i$, and the veiling factor  is $r_\lambda(i)=F_\lambda^{\mathrm{warm}}(i)/F_\lambda^{\mathrm{hot}}(i)$.

Assuming that the intrinsic absorption of gas is identical in both phases ($\tilde{F}_{\lambda,o}(B)=\tilde{F}_{\lambda,o}(Q)$), the burst-phase profile can be written in terms of the quiescent-phase profile:
\begin{equation}
\tilde{F}_\lambda(B) =\left(\frac{1+r_\lambda(Q)}{1+r_\lambda(B)}\right)\left(\tilde{F}_\lambda(Q) + \frac{r_\lambda(B)-r_\lambda(Q)}{1+r_\lambda(Q)} \right).
\label{eq:v1}
\end{equation}
Introducing the relative veiling, $\mathcal{V}$:
\begin{equation}
1+\mathcal{V}
\equiv\frac{1+r_\lambda(B)}{1+r_\lambda(Q)}
=\left(\frac{F_\lambda^{\mathrm{cont}}(B)}{F_\lambda^{\mathrm{cont}}(Q)}\right)\big/\left(\frac{F_\lambda^{\mathrm{hot}}(B)}{F_\lambda^{\mathrm{hot}}(Q)}\right),
\end{equation}
with $F_\lambda^{\mathrm{cont}}(i)=F_\lambda^{\mathrm{warm}}(i)+F_\lambda^{\mathrm{hot}}(i)$,
Equation~(\ref{eq:v1}) reduces to:
\begin{equation}
\tilde{F}_\lambda(B) =\frac{\tilde{F}_\lambda(Q) + \mathcal{V}}{1+\mathcal{V}}.
\label{eq:veiling_ratio}
\end{equation}
This expression is more general than Equation~(\ref{eq:veiling}) and applies even when both epochs experience nonzero veiling. In the limiting case that the quiescent spectrum is unveiled, $r_\lambda(Q)=0$, the relative veiling reduces to the classical veiling in the burst epoch, $\mathcal{V}_\lambda \rightarrow r_\lambda(B)$, and Equation~(\ref{eq:veiling_ratio}) becomes equivalent to Equation~(\ref{eq:veiling}).

\subsection{Hot continuum emission sources}\label{subsec:method_hot}
A main goal of this study is to relate the variability of the hot continuum that backlights the CO and \water\ absorption lines to changes in the mass accretion rate through the inner disk (see Section~\ref{sec:intro}). To do so, we require a physical model that links the continuum flux at a given wavelength to the temperature structure and radial extent of the emitting region. Throughout this work we interpret the hot midplane continuum primarily in terms of a viscously heated, optically thick inner disk. In a steady viscous disk, the midplane temperature profile is determined by the mass accretion rate $\dot{M}$ and the inner truncation radius.
In this subsection we summarize the temperature structure of a viscously heated disk and the expected spectrum of the magnetospheric accretion shock, which together provide the mapping between the hot midplane continuum emission and $\dot{M}$. The specific constraints on $\dot{M}$ in EC 53 derived from this framework are presented in Section~\ref{subsec:analysis_mdot}, with broader implications discussed in Section~\ref{subsec:dis_mdot}.

A viscously heated disk is a plausible candidate for the source of the hot midplane continuum emission \citep[e.g.,][]{Labdon2021,Yoon2021}.
The temperature profile of the viscously heated disk is given by
\begin{equation} \label{eq:T_vis}
    T^4_{\rm visc}(R) = \frac{3GM_*\dot{M}}{8\pi\sigma R^3}\left( 1-\sqrt{\frac{R_{\rm in}}{R}}\right),
\end{equation}
where $M_*$ is the stellar mass, $\dot{M}$ is the mass accretion rate, $\sigma$ is the Stefan-Boltzmann constant, and $R_{\rm in}$ is the inner boundary of the disk \citep[e.g.,][]{Popham1996,Liu2022}. 

The innermost gaseous disk ($R_{\rm in}$) may be truncated by magnetic field \citep{Bouvier2007,Hartmann2016}:
\begin{eqnarray}\label{eq:r_in}
     R_{\mathrm in} &\simeq& 1.4\,R_* \left( \frac{B_*}{1\,\mathrm{kG}} \right)^{4/7} \left( \frac{R_*}{1.6\,R_\odot} \right)^{5/7} \\ \nonumber
     & & \times \left( \frac{M_*}{0.3\,M_\odot} \right)^{-1/7} \left( \frac{\dot M}{2.0 \times10^{-6}\,M_\odot\,\mathrm{yr}^{-1}}\right)^{-2/7}.
\end{eqnarray}
where $B_*$ is the stellar magnetic field strength, and $R_*$ is the stellar radius.  

The emergent hot midplane continuum emission from the region hotter than a given temperature $T$ is computed as:
\begin{equation} \label{eq:flux_vis}
    F_\lambda d\lambda = \int^{R(T)}_{R_{\rm in}} B_{\lambda}(T_{\rm visc}) 2\pi R dR d\lambda *\cos(i)/(\pi d_*^2)
\end{equation}
where $B_{\lambda}$ is the Planck function, $d_*$ is the distance to EC 53, and $i$ is the inclination. 

Magnetospheric accretion shocks primarily contribute at UV and optical wavelengths, and may provide a minor contribution to the infrared continuum in systems with high mass accretion rates, such as EC 53 \citep[e.g.,][]{Calvet1998}. Its contribution can be approximated following \citet{Liu2022}:
\begin{equation}\label{eq:flux_mag}
    F_\lambda d\lambda = B_{\lambda}(T_{\rm mag})d\lambda\times \frac{GM_*\dot{M}}{R_*}\left(1 - \frac{R_*}{R_{\rm in}}\right)/(\sigma T_{\rm mag}^4)/(\pi d_*^2),
\end{equation}
where the shock-heated region is assumed to emit as a blackbody with an effective temperature of $T_{\rm mag}$= 8000\,K.

Note that the viscous-heating prescription adopted here (Equation~\ref{eq:T_vis}) is derived for a steady, conservative disk with a radially constant accretion rate, $\dot{M}(r)={\rm const}$ \citep[e.g.,][]{Pringle1981}. In more general cases where the instantaneous disk accretion rate varies with radius (e.g., due to time-dependent mass accumulation/release and/or disk winds that remove mass and/or angular momentum), the local dissipation and temperature structure are correspondingly modified and the mapping between $\dot M$ and $T(r)$ becomes more complex \citep[see, e.g.,][]{Knigge1999}. In this work, we therefore use the standard constant-$\dot{M}$ form as a simplified, phenomenological prescription to relate the inferred hot-continuum emission (and its change between epochs) to a representative accretion rate characterizing the disk region that produces the backlighting continuum for a given molecular band.

We also clarify our notation. In Equation~(\ref{eq:T_vis}), $\dot M$ denotes the mass accretion rate \emph{through the disk} ($\dot{M}_{\rm disk}$), whereas in Equations~(\ref{eq:r_in}) and~(\ref{eq:flux_mag}) it denotes the mass accretion rate \emph{onto the protostar} ($\dot{M}_*$). For simplicity, throughout the analysis in Sections~4--5 we evaluate these expressions assuming a radially constant accretion rate, $\dot{M}_*=\dot{M}_{\rm disk}\equiv\dot M$. In Section~\ref{subsec:dis_mdot}, we relax this assumption in a toy framework by allowing $\dot M_{\rm disk}$ to vary with radius, and we distinguish it from the accretion rate onto the protostar, $\dot M_*$.

\section{Analysis} \label{sec:analysis}
In this section, we apply the analysis framework developed in Section~\ref{sec:methods} to derive the burst-to-quiescent mass-accretion-rate ratio, $\dot{M}(B)/\dot{M}(Q)$. First, we use LTE slab models to determine the gas temperatures and column densities of the absorbing gas in each phase (Section~\ref{subsec:analysis_lte}). 
We then quantify the line-depth variations between epochs with the \emph{relative veiling}, $\mathcal{V}$ (Section~\ref{subsec:analysis_veiling ratio}). Combining the observed continuum emission ratios, $F_\lambda^{\mathrm{cont}}(B)/F_\lambda^{\mathrm{cont}}(Q)$, with the derived relative veiling values, we infer the corresponding hot midplane continuum emission ratios, $F_\lambda^{\mathrm{hot}}(B)/F_\lambda^{\mathrm{hot}}(Q)$ (Section~\ref{subsec:analysis_hot}). Finally, we compare these continuum emission ratios with those predicted by a viscously heated disk model to constrain the mass-accretion-rate ratio between the two phases (Section~\ref{subsec:analysis_mdot}).

\subsection{LTE models of CO and \water\ absorption lines} \label{subsec:analysis_lte}
As mentioned in Section~\ref{subsec:method_lte}, LTE slab models provide estimates of gas temperatures and column densities for the absorbing gas, and thereby a lower bound on the temperature of the hot midplane continuum source. However, these estimates are subject to large uncertainties due to the simplifying assumptions of the model, including the use of a single-temperature LTE component, uncertainties in the continuum level, and potential degeneracies between temperature and column density. In addition, the true uncertainties are difficult to quantify robustly, as they depend on model assumptions and parameter degeneracies that are not fully captured by formal fitting errors. Therefore, the derived parameters should be interpreted as approximate values rather than precise measurements. Here we apply these models to the IGRINS and JWST spectra to derive the gas temperatures and column densities in both epochs, with a focus on relative comparisons between the burst and quiescent phases.

For the IGRINS data, a hot \twco\ slab with $T_{\rm gas}\simeq1800$\,K and $N(\twco)\sim10^{20}$\,cm$^{-2}$ provides the best fit to the broad overtone absorption (Table~\ref{tb:co_overtone_model}; pink model in Figure~\ref{fig:co_overtone_igrins}). The inferred CO column density is consistent with values reported for hot inner disk gas traced by CO overtone emission in previous studies \citep[e.g.,][]{SLee2016,Contreras2017,Kraus2000}. However, given the simplifying assumptions of the slab model---in particular the adoption of a relatively broad intrinsic line width and the assumption that the background continuum is significantly hotter than the absorbing gas, both of which act to reduce the required column density---this value could be regarded as a lower limit on the true CO column density.  The \water\ stretching mode, although limited by a relatively low SNR, is independently fitted in both temperature and column density. The resulting best-fit temperature is consistent with that of the \twco\ component within the uncertainties.

\begin{figure*}[htp]
 \centering
 \vspace{-2mm}
 \includegraphics[width=1.0\textwidth]{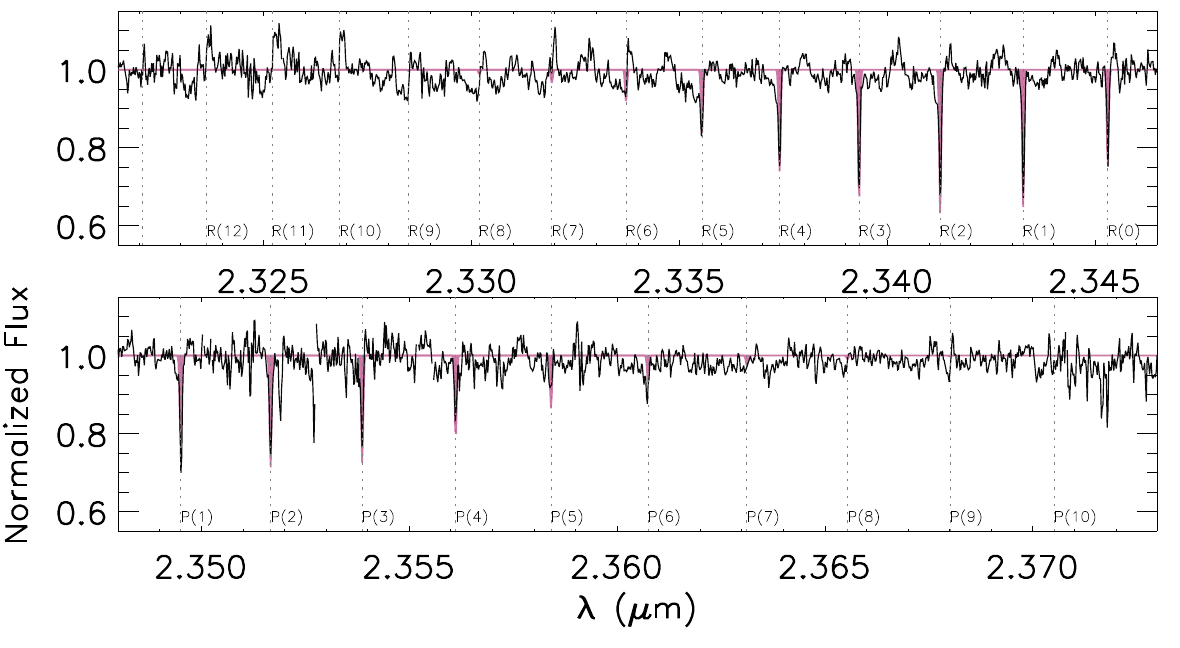}
  \vspace{-5mm}
 \caption{ Residual spectrum from Figure~\ref{fig:co_overtone_igrins} (black), obtained after subtracting the hot  components, together with the best-fit cold CO model (pink). This zoomed-in view focuses on the R(7)–P(6) region to highlight the narrow absorption component. The model parameters are listed in Table~\ref{tb:co_overtone_model}. The vertical dotted lines indicate the wavelengths of \twco\ $\nu$=2--0 transitions.}
\label{fig:co_overtone_igrins_res}
\end{figure*}

The single hot-component model, however, does not reproduce the low-$J$ $v=2$–0 lines from R(5) to P(5), which display additional narrow absorption components (Section~\ref{subsec:resultsIgrins}). As shown in Figure~\ref{fig:co_overtone_igrins_res}, the residual spectrum reveals narrow absorption features that are clearly distinguished from the broad component over this limited range of transitions. 
To account for these features, we add a second, narrow \twco\ slab with an assumed line width of 1\,\kms. A model with $T_{\rm gas}\simeq23$\,K and $N(\twco)\simeq4\times10^{19}$\,cm$^{-2}$ reproduces the narrow overtone absorption remarkably well.  At higher rotational levels, weak residual structures are present, but it is unclear whether these arise from additional physical components or from calibration uncertainties and limitations in the data. We therefore refrain from further interpretation of these features. 

We next model the CO fundamental band and associated \water\ lines in the JWST spectra. We adopt a Gaussian line width consistent with that used in the IGRINS-based modeling, without explicitly modeling the underlying kinematic structure. 
In the quiescent phase, the CO fundamental absorption is broadly reproduced with a temperature of $T_{\mathrm{gas}}\simeq1130$\,K and a column density of $N(\twco)\simeq3.03\times10^{18}$\,cm$^{-2}$ (Table~\ref{tb:jwst_model}; Figure~\ref{fig:co_fudamental_obs_model}). However, significant residuals remain in the 4.6–4.7\,\micron\ region, indicating that the single-temperature model does not fully capture the observed line structure. These residuals may arise from a combination of continuum uncertainties (e.g., partial overlap with CO ice absorption) and additional structure in the warm CO gas, suggesting that a single-temperature description is insufficient. 
Assuming the same kinetic temperature for \thco\ and \water\ as for \twco, we obtain $N(\thco)\simeq1.5\times10^{17}$\,cm$^{-2}$ and $N(\mathrm{H_2O})\simeq3.2\times10^{18}$\,cm$^{-2}$ in the CO fundamental region. 

In the burst phase, the CO fundamental absorption lines are systematically shallower than in the quiescent phase, while the relative strengths among the $v=1$--0 transitions remain largely unchanged (Figure~\ref{fig:co_fudamental_obs_model}). Because the relative line strengths primarily trace the gas temperature, this similarity indicates that the gas temperature does not change substantially between the two phases. In contrast, the reduced absolute line depths suggest a lower apparent CO column density during the burst phase. Consistent with this qualitative behavior,  the LTE slab fits favor a modest increase in the best-fit temperature by $\sim200$\,K and a decrease in the $^{12}$CO column density by a factor of $\sim2.6$ in the burst phase.  The fitted \thco\ and \water\ column densities in the burst are also lower than in the quiescent phase (Table~\ref{tb:jwst_model}), suggesting an apparent reduction in the amount of warm molecular gas along the line of sight.

As shown in Figures~\ref{fig:h2o_2.7um} and \ref{fig:h2o_bending_obs_model}, a single-temperature LTE component does not capture all the detailed structure of the \water\ stretching and bending mode profiles. Nevertheless, within the framework of the single-temperature LTE model, the inferred \water\ column densities provide a consistent basis for comparing the two epochs, although their absolute values should be interpreted with caution. In the stretching mode, we adopt the gas temperature derived from the IGRINS 2.3\,\um\ \water\ fitting and apply it to both epochs, while for the bending mode we adopt the temperature inferred from the CO fundamental analysis. In the stretching mode, the columns in the burst and quiescent phases agree within the uncertainties, so the amount of hot \water\ traced by this band shows little or no variability. In contrast, the \water\ bending mode absorption observed with MIRI (4.9–7.6\,\micron) exhibits a decrease in column density by about 30\% from quiescence to burst, in line with the weaker bending mode absorption in the burst phase spectra (Figure~\ref{fig:h2o_bending_obs_model}). 

However, given that the warm region expands during the burst phase, a decrease in the column density of the gas responsible for the CO fundamental band and \water\ bending mode features is not a natural expectation (see Section~\ref{subsec:discuss_cofund}).

We note that the uncertainties derived from the LTE slab fitting represent conditional statistical errors only. These uncertainties are estimated using a bootstrap approach in which each parameter is varied while the others are held fixed at their best-fit values, and the resulting standard deviation is adopted as the 1$\sigma$ uncertainty. As a result, the reported values reflect the local sensitivity of the model to individual parameters and do not capture parameter covariances or additional systematic uncertainties. The true uncertainties are therefore likely larger than the formal values reported in Tables~\ref{tb:co_overtone_model} and \ref{tb:jwst_model}.
\begin{deluxetable}{lcc}
\tabletypesize{\footnotesize}   
\tablecaption{Best-fit IGRINS Model Parameters\label{tb:co_overtone_model}}
\tablehead{
  \colhead{Mol.} &
  \colhead{$T_{\rm gas}$ (K)} &
  \colhead{$N$ (cm$^{-2}$)}}
\startdata
\twco\          & 1780\,(68) & $1.04\,(0.04)\times10^{20}$\\
\twco\ (cold)   &   23\,(10)   & $3.6\,(1.9)\times10^{19}$\\
\water             & 1870\,(151)  & $0.99\,(0.16)\times10^{20}$ \\
\enddata
\tablecomments{Values in parentheses give the $1\sigma$ uncertainties, representing conditional statistical errors estimated  with the other slab parameters held fixed.}
\end{deluxetable}

\begin{deluxetable*}{cccc}
\tablecaption{Model parameters \label{tb:jwst_model}}
\tablehead{
\colhead{Parameters} & \colhead{Quiescence} & \colhead{Burst} &\colhead{Burst($\mathcal{V}$)$^a$}} 
\startdata
NIRSpec (2.3\,\um)\\
\hline
 $N(\mathrm{H_2O})^b$ ($\times10^{18}$ cm$^{-2}$) & 5.70 (0.20) & 6.21 (0.25) & -\\
\hline 
NIRSpec (4.6\,\um)\\
\hline 
 $T_{\rm gas}$:  CO (K)                            & 1130 (14)   & 1340 (32)   & 1190 (32)\\
 $N(^{12}\mathrm{CO})$ ($\times10^{18}$ cm$^{-2}$) & 3.03 (0.004) & 1.18 (0.004) & 3.22 (0.013)\\
 $N(^{13}\mathrm{CO})^c$ ($\times10^{17}$ cm$^{-2}$)  & 1.49 (0.01) & 0.90 (0.01) & 1.58 (0.02)\\
 $N(\mathrm{H_2O})^c$ ($\times10^{18}$ cm$^{-2}$)     & 3.18 (0.03) & 2.86 (0.03) & 5.09 (0.06)\\
 \hline
 MIRI \\
 \hline
 $N(\mathrm{H_2O})^c$ ($\times10^{18}$ cm$^{-2}$)    & 1.10 (0.06) & 0.7 (0.05) & 1.25 (0.09) 
\enddata
\tablecomments{Values in parentheses give the $1\sigma$ uncertainties, representing conditional statistical errors estimated  with the other slab parameters held fixed.\\
$^a$ The relative veilings ($\mathcal{V}$) are 0.906, 0.83 for the CO fundamental band and  \water\ bending mode, respectively.\\
$^b$The temperature is adopted from Table~\ref{tb:co_overtone_model} (1800\,K).  \\
$^c$The temperatures are assumed to be the same as those of \twco.  }
\end{deluxetable*}

\begin{figure*}[htp]
 \centering
 \vspace{-2mm}
 \includegraphics[width=1.0\textwidth]{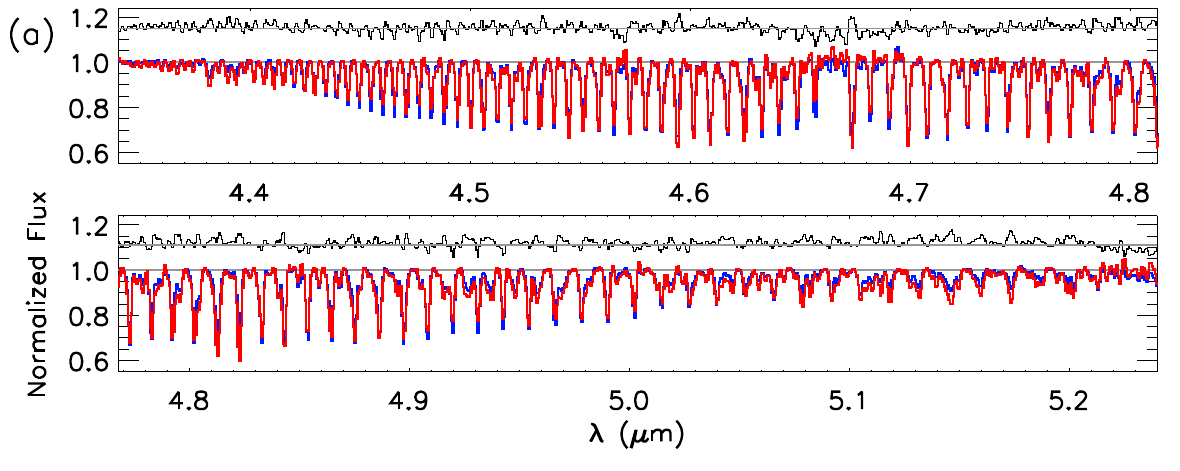}
 \includegraphics[width=1.0\textwidth]{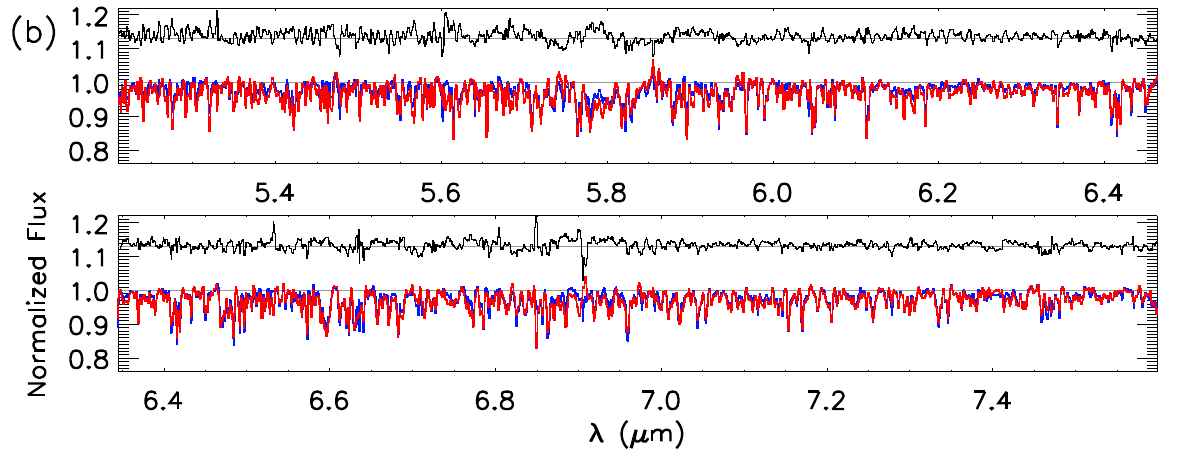}
 \vspace{-5mm}
 \caption{(a) Normalized CO fundamental band absorption spectra during the burst (red) and quiescent (blue) phases. The burst phase spectrum has been adjusted using a relative veiling factor of $\mathcal{V}=0.906$, derived as described in Section~\ref{subsec:analysis_veiling ratio}. The black spectrum shown above the data represents the difference between the quiescent spectrum and the  adjusted burst spectrum, offset vertically for clarity.
(b) Same as panel (a), but for the \water\ bending mode absorption observed with MIRI, where a relative veiling factor of $\mathcal{V}=0.83$ is applied.}
\label{fig:co_h2o_veiling}
\end{figure*}

\subsection{Relative veiling\,($\mathcal{V}$)} \label{subsec:analysis_veiling ratio}
In this section, we consider an alternative explanation in which the variations observed in the CO fundamental band and the \water\ bending mode are primarily driven by changes in the continuum emission. To test this continuum-dilution scenario, we compare the two epochs using the relative veiling formalism introduced in Section~\ref{subsec:method_veiling_ratio}. Figure~\ref{fig:co_h2o_veiling} illustrates, for both species, the difference between the burst spectrum and the quiescent spectrum after applying the relative veiling correction to the burst phase. The residuals are minimized when the burst CO fundamental and \water\ bending–mode spectra are corrected with relative veilings of $\mathcal{V}=0.906\pm0.004$ and $\mathcal{V}=0.83\pm0.01$, respectively, indicating that these values yield the closest match between the two epochs.

As a consistency check, we repeated the LTE slab fits from Section~\ref{subsec:analysis_lte} after adjusting the burst-phase spectra based on the inferred relative veiling values. The resulting temperatures and column densities (Table~\ref{tb:jwst_model}) are nearly identical between the two epochs. Thus, after applying the relative veiling correction, the apparent differences in the LTE-derived gas properties between the two epochs largely disappear, yielding gas temperatures and column densities that are effectively consistent within the uncertainties.

\subsection{Variation in the hot midplane continuum emission} \label{subsec:analysis_hot}
The ratio of the hot midplane continuum emission between the two phases, $F_\lambda^{\mathrm{hot}}(B)/F_\lambda^{\mathrm{hot}}(Q)$, is estimated using the following relation:
\begin{equation}\label{eq:f_hot}
    \frac{F_\lambda^{\mathrm{hot}}(B)}{F_\lambda^{\mathrm{hot}}(Q)} =  \frac{F_\lambda^{\mathrm{cont}}(B)}{F_\lambda^{\mathrm{cont}}(Q)}\frac{1}{1 + \mathcal{V}}.
\end{equation}
Note that $F_\lambda^{\mathrm{cont}}(B)/F_\lambda^{\mathrm{cont}}(Q)$ can be replaced by the observed ratio of the total continuum emission because the total continuum emission observed in the burst and quiescent phases suffers the same extinction from the envelope and foreground materials.

For the CO overtone and \water\ stretching mode lines, the relative veiling $\mathcal{V}$ is negligible, so the observed continuum emission ratios directly approximate the hot midplane continuum emission ratios, yielding $F_\lambda^{\rm hot}(B)/F_\lambda^{\rm hot}(Q)=2.9\pm0.3$ and $3.0\pm0.2$, respectively (see Figure~\ref{fig:jwst_zoom}). In contrast, for the CO fundamental and \water\ bending mode rovibrational lines, the hot midplane continuum emission ratios, derived by using Equation~(\ref{eq:f_hot}) and the parameters listed in Table~\ref{tb:jwst_model}, are $F_\lambda^{\mathrm{hot}}(B)/F_\lambda^{\mathrm{hot}}(Q)$=1.71$\pm$0.04 and  1.9$\pm$0.1, respectively.

\begin{figure*}[htp]
 \centering
 \vspace{-2mm}
 \includegraphics[width=0.8\textwidth]{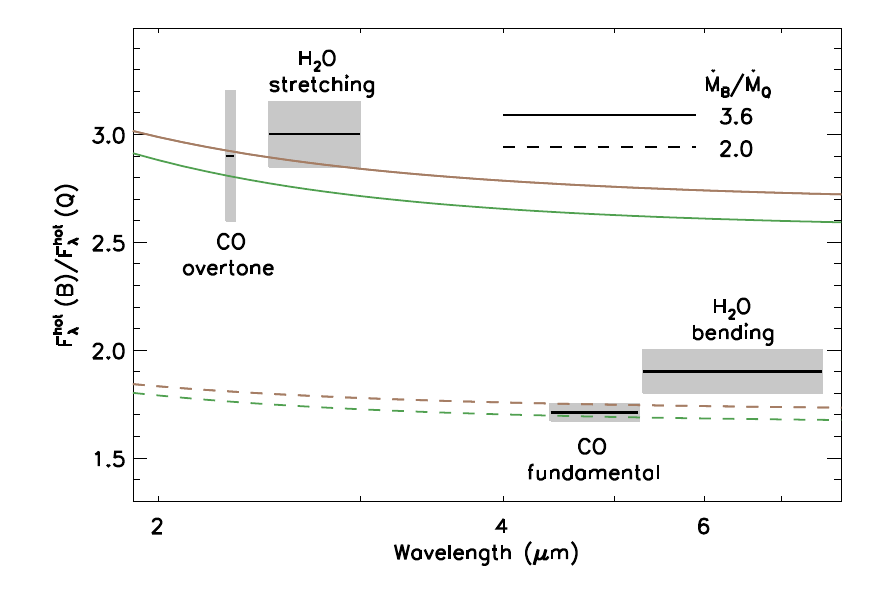}
 \vspace{-5mm}
 \caption{Comparison of observed and model hot midplane continuum emission ratios between burst (B) and quiescent (Q) phases. Black bars with 1$\sigma$ error bars show the $F_\lambda^{\mathrm{hot}}(B)/F_\lambda^{\mathrm{hot}}(Q)$ measured from the CO overtone, \water\ stretching mode, CO fundamental, and \water\ bending mode absorption lines (see Section~\ref{subsec:analysis_hot}). Colored curves show viscous‐disk model predictions integrated over regions where the midplane temperature exceeds 1800\,K (brown) and 1200\,K (green). Solid and dashed curves correspond to burst‐to‐quiescent mass-accretion-rate ratios of 3.6 and 2.0, respectively.}
 \label{fig:liu_model}
\end{figure*}

\begin{deluxetable*}{cccccc}
\tablecaption{Ratios between the burst and quiescent phases\label{tb:ratios}}
\tablehead{
    \colhead{\bf Transitions} & \colhead{\bf $T_{\rm gas} (K)$} & \colhead{\bf $F_\lambda^{\mathrm{cont}}(B)/F_\lambda^{\mathrm{cont}}(Q)$} & \colhead{$\mathcal{V}$} & \colhead{\bf $F_\lambda^{\rm hot}(B)/F_\lambda^{\rm hot}(Q)$}  & \colhead{\bf $\dot{M}(B)/\dot{M}(Q)$} }
\startdata
 CO overtone band  (2.3 \um)       & 1780 (68)                &  2.88 (0.01)  & $\sim$0 (0.1)   & 2.9 (0.3)   & $\sim$3.6 (0.5)\\
 \water\ stretching mode (2.8 \um) & - & 3.0 (0.03)     & $\sim$0 (0.05)   & 3.00 (0.15)   &$\sim$3.8 (0.3)\\
  CO fundamental band (4.6 \um)     &1130 (14) /1190 (32)     & 3.25 (0.07)    & 0.906 (0.004) & 1.71 (0.04) &$\sim$2.0 (0.1)\\
\water\ bending mode ( 6 \um)      & - &                       3.5 (0.1)      & 0.83 (0.01)  & 1.9 (0.1)  & $\sim$2.4 (0.2) 
 \enddata
 \tablecomments{Columns are as follows:
(2) gas temperature from the LTE model (see Section \ref{subsec:analysis_lte});
(3) observed continuum emission ratio;
(4) relative veiling (see Section \ref{subsec:analysis_veiling ratio});
(5) hot midplane continuum emission ratio (see Section \ref{subsec:analysis_hot});
(6) mass-accretion-rate ratio (see Section \ref{subsec:analysis_mdot}).\\
 The values in parentheses are 1 $\sigma$ uncertainties; those for the CO overtone band and  \water\ stretching mode are $\sim$ 1/(SNR of absorption lines).}
\end{deluxetable*}
\subsection{Variation in mass-accretion-rate ratio}\label{subsec:analysis_mdot}
We next translate the inferred hot-continuum ratios into constraints on the burst-to-quiescent mass-accretion-rate ratio, $\dot{M}(B)/\dot{M}(Q)$, using a simple viscous-heating prescription for the inner disk.  
The hot-continuum ratios themselves (Table~\ref{tb:ratios}, Columns 3--5) are primarily constrained by the observations and include formal error propagation, although additional systematic uncertainties may arise from continuum estimation and instrumental effects.  
The conversion of these ratios into a mass-accretion-rate ratio is more model-dependent. However, because this analysis is based on an epoch-to-epoch continuum ratio rather than on an absolute continuum luminosity, its sensitivity to the uncertain stellar parameters is reduced. In Equation~(\ref{eq:flux_vis}), the inclination enters only as a multiplicative $\cos i$ term and therefore cancels in $F_{\lambda}^{\rm hot}(B)/F_{\lambda}^{\rm hot}(Q)$. In addition, once the quiescent accretion luminosity is used as an empirical anchor, the dependence on the stellar mass is substantially reduced. 

The remaining stellar-parameter sensitivity enters mainly through the truncation radius, and is therefore driven primarily by the adopted stellar radius and magnetic-field strength.  
We adopt $M_*=0.3\,M_\odot$ and a quiescent-phase accretion rate of $\dot{M}(Q)=2.0\times10^{-6}\,M_\odot\,{\rm yr^{-1}}$ \citep{Baek2020,yhLee2020}. With $B_*=1$\,kG \citep{Johns-Krull2007,Donati2009,Flores2024} and $R_*=1.6\,R_\odot$, Equation~(\ref{eq:r_in}) yields a truncation radius of $R_{\rm in}=1.4\,R_*$ (0.01\,au) in quiescence.  
We therefore regard the inferred accretion-rate ratio as a representative, model-dependent estimate that tests whether the observed continuum variations are broadly consistent with plausible accretion-rate changes, rather than as a uniquely determined value.

For the quiescent disk, we compute the viscous-heating temperature profile $T_{\rm vis}(r)$ (Equation~\ref{eq:T_vis}) and identify the radial extents of the midplane zones hotter than two representative thresholds, $T>1800$\,K and $T>1200$\,K (extending to $\sim$0.05\,au and $\sim$0.1\,au, respectively). We then use Equation~(\ref{eq:flux_vis}) to integrate the continuum emission from each hot zone and obtain the corresponding quiescent-phase hot-continuum fluxes, $F_{\lambda}^{\rm hot}(Q)$.

To model the burst phase, we treat the accretion-rate enhancement $\dot{M}(B)/\dot{M}(Q)$ as a free parameter. For each trial value, we recompute $R_{\rm in}$ and $T_{\rm vis}(r)$ using the scaled accretion rate $\dot{M}(B)$, re-identify the regions satisfying $T>1800$\,K and $T>1200$\,K, and evaluate the corresponding hot-zone continuum fluxes, $F_{\lambda}^{\rm hot}(B)$, via Equation~(\ref{eq:flux_vis}). This procedure yields the predicted hot-continuum ratio $F_{\lambda}^{\rm hot}(B)/F_{\lambda}^{\rm hot}(Q)$ as a function of $\dot{M}(B)/\dot{M}(Q)$ for each temperature threshold, producing the model curves shown in Figure~\ref{fig:liu_model}.

Figure~\ref{fig:liu_model} summarizes these predictions: the brown and green curves correspond to continuum emission from midplane regions with $T>1800$\,K and $T>1200$\,K, respectively. Because the CO overtone and fundamental rovibrational absorption arises in the disk atmosphere and traces gas at characteristic temperatures of $\sim$1800\,K and $\sim$1200\,K, respectively, we adopt these values as fiducial lower limits for the temperature of the backlighting hot-continuum component. Accordingly, we compare the CO overtone band (and the \water\ stretching mode) to the $T>1800$\,K curves, and the CO fundamental band (and the \water\ bending mode) to the $T>1200$\,K curves. The solid and dashed line styles illustrate burst-to-quiescent accretion-rate ratios of $\dot{M}(B)/\dot{M}(Q)=3.6$ and 2.0, respectively.

For the CO overtone band, the inferred hot-continuum ratio $F_\lambda^{\mathrm{hot}}(B)/F_\lambda^{\mathrm{hot}}(Q)=2.9\pm0.3$ is reproduced by $\dot{M}(B)/\dot{M}(Q)=3.6\pm0.5$ (using the $T>1800$\,K curves). For the CO fundamental band, the observed $F_\lambda^{\mathrm{hot}}(B)/F_\lambda^{\mathrm{hot}}(Q)=1.71\pm0.04$ is reproduced by a lower accretion-rate ratio of $\dot{M}(B)/\dot{M}(Q)\approx2.0$ (using the $T>1200$\,K curves).

The \water\ rovibrational absorption shows behavior closely analogous to that of CO. The IGRINS spectra provide an independent constraint on the \water\ stretching-mode temperature, indicating that it traces gas at temperatures comparable to those of the CO overtone band; the best-fit temperatures are consistent within the uncertainties. By contrast, the temperature of the \water\ bending mode is not independently constrained in our JWST fits and is therefore assumed to be the same as that of the CO fundamental band, motivated by its association with cooler gas at larger radii. Within this framework, the \water\ stretching and bending modes probe regions analogous to those traced by the CO overtone and fundamental transitions, respectively. The corresponding $\dot{M}(\mathrm{B})/\dot{M}(\mathrm{Q})$ values inferred from the two \water\ modes are summarized in Table~\ref{tb:ratios}.

We note that an additional source of model dependence in this analysis is the choice of the temperature boundary used to define the hot- and warm-midplanes. Here we adopt the gas temperatures inferred from the slab models as physically motivated lower limits. If the effective continuum temperature thresholds are higher than these adopted values, the representative accretion-rate ratios derived here would change accordingly. We therefore interpret them as band-weighted, phenomenological estimates rather than as a unique measurement of a radial accretion-rate profile.
\section{Discussions}\label{sec:discussion}

\subsection{Assessing the Origin of the CO Overtone Absorption}

The origin of the CO overtone absorption provides an important constraint on the physical interpretation of the two-layer model adopted in this work. In particular, the high spectral resolution of the IGRINS data allows us to resolve the line profiles and thus place meaningful constraints on the kinematic origin of the absorbing gas, complementing the JWST observations.

In our framework, the CO overtone absorption traces hot gas with a temperature of $\sim$1800\,K located above the hot midplane. Based on the viscous-heating model, this temperature corresponds to a radius of $\sim$0.09\,au during the burst phase. At this radius, the projected Keplerian velocity is $\sim$31\,\kms\ for the adopted inclination of EC\,53 ($i \simeq 34.5^\circ$; \citealt{sLee2020}). Because a Keplerian line profile reflects the full range of projected velocities across the disk, this velocity scale would typically produce a broader and, depending on the disk geometry, potentially double-peaked profile than observed. The observed single-peaked line profile and relatively narrow width (FWHM $\sim$25\,\kms) therefore indicate that pure Keplerian rotation alone cannot fully explain the line shape.

The absorbing gas is unlikely to be associated with the outflowing material identified in Paper III. That outflow is observed at larger distances from the central source and exhibits significantly higher velocities ($>$10–60\,\kms), whereas the CO overtone absorption is centered close to the systemic velocity, with only a small blueshift ($\sim$-3\,\kms). In addition, CO absorption associated with outflows in embedded protostars is often characterized by strongly blueshifted and asymmetric profiles, sometimes showing discrete velocity components \citep[e.g.,][]{Herczeg2011}, in contrast to the approximately symmetric and centrally peaked profile observed in EC\,53. These differences suggest that the CO overtone absorption in EC\,53 is not dominated by the large-scale outflow component.

The high temperature, large CO column density, proximity to the systemic velocity, and absorption geometry collectively suggest that an origin in the inner disk is the most plausible interpretation, in which the hot midplane provides the background continuum and the CO gas resides in the disk atmosphere. This configuration is consistent with the adopted two-layer model, although it cannot be uniquely established.

The observed line profile can be more naturally explained if the disk atmosphere is not purely Keplerian, but includes sub-Keplerian kinematics, such as those expected in disk winds. In this picture, the absorbing gas remains physically associated with the inner disk atmosphere, while the reduced projected velocities and single-peaked profiles arise from a combination of rotational and vertical motions \citep[e.g.,][]{Pontoppidan2011,Banzatti2022}. Thus, the disk wind could be regarded as a modification of the disk atmosphere kinematics rather than a distinct spatial component.

Compared to more evolved Class~II systems, EC\,53 is characterized by a higher mass accretion rate and a more deeply embedded environment, which likely enhances the inner disk continuum and allows the hot disk atmosphere to be observed in absorption against the bright background. In this sense, the CO overtone absorption in EC\,53 may represent the embedded-phase counterpart of molecular gas that is more commonly observed in emission in less embedded disks \citep[e.g.,][]{Herczeg2011,Brown2013}.

While the inner disk atmosphere provides the most natural origin for the CO overtone absorption, the detailed kinematic structure of the gas remains uncertain. In particular, the relative contributions of Keplerian rotation and sub-Keplerian motions cannot be uniquely constrained, and a minor contribution from slow-moving outflowing material cannot be entirely excluded.

\subsection{Assessing the Origin of the CO Fundamental Absorption Variability}
\label{subsec:discuss_cofund}

In this section, we investigate the origin of the reduced normalized depth of the CO fundamental rovibrational absorption during the burst relative to quiescence. We consider four scenarios and evaluate each against the constraints provided by the isotopologue behavior, the correlated multi-band absorption trends (CO and \water), and the lack of large-amplitude variability in the emission features within the same extraction aperture.

\subsubsection{Scenario 1: A genuine decrease in the CO column density}
\label{subsubsec:scen_col}

A straightforward interpretation is that the line-of-sight CO column density decreases during the burst. However, this explanation is disfavored by the relative behavior of $^{12}$CO and $^{13}$CO. Because the $^{12}$CO fundamental transitions are expected to be more optically thick than the corresponding $^{13}$CO lines, a substantial intrinsic reduction in the absorbing column would generally produce a larger fractional change in the more optically thin $^{13}$CO absorption than in $^{12}$CO. Instead, Table~\ref{tb:jwst_model}  indicates that the inferred column-density change is larger for $^{12}$CO than for $^{13}$CO, arguing against a scenario in which the observed variability is driven primarily by a true decrease in the CO column density.

This interpretation is further supported by the expected chemical resilience of CO in the inner gaseous disk. CO is expected to remain abundant in the inner gaseous disk up to its thermal dissociation temperature of $\sim$4000\,K \citep{Najita2007}, and its high formation rate can efficiently compensate for photodissociation \citep{Thi2005}. We therefore consider a genuine decrease in the CO column density to be an unlikely primary driver of the observed variability.

\subsubsection{Scenario 2: Changes in the effective $T_{\rm bg}/T_{\rm gas}$ ratio in the optically thick limit}
\label{subsubsec:scen_tbg}

The apparent variability may arise from changes in the effective ratio between the absorbing-gas temperature and the background brightness temperature in the optically thick limit. In this case, the normalized line-core flux approaches the ratio of the line source function to the background continuum brightness, so the absorption becomes shallower as the background continuum brightness temperature approaches the absorbing-gas temperature.

For our LTE slab fits, the gas temperatures are $1130\pm14$ K and $1340\pm32$ K for the quiescent and burst phases, respectively. During quiescence, the CO fundamental line cores reach a normalized depth of $\approx0.3$, whereas during the burst they are shallower, with a depth of $\approx0.15$. This factor-of-two reduction in depth is consistent with an effective background continuum brightness temperature of $\sim1500$ K at 4.6\,\um.

However, the relatively shallow absolute depths indicate that an additional dilution of the line-to-continuum contrast is still required within this framework. A simple parameterization suggests that an order-unity veiling at 4.6\,\um\ is needed; adopting the observed depths implies $r_{4.6\,\mu m}\approx0.8$ in both epochs to reproduce the absolute line-to-continuum contrast. While this scenario demonstrates that changes in the relative brightness of the backlighting continuum compared to the line source function can in principle contribute to the apparent variability, applying this interpretation in a physically self-consistent manner requires a more detailed inner-disk model that couples the radial temperature structure, emitting area, and radiative transfer (including optical depth and covering fraction). We therefore treat this scenario as a plausible contributing effect rather than adopting it as the primary explanation.

\subsubsection{Scenario 3: Emission filling by unresolved CO emission within the extraction aperture}
\label{subsubsec:scen_fill}

A third possibility is that the apparent weakening of the CO fundamental absorption arises from an enhanced CO emission component that we cannot isolate at our current spectral resolution, such that emission filling reduces the normalized absorption depth. Paper~III demonstrates that CO emission is present in the outflow cavity of EC~53, and therefore we cannot ignore the possibility that an unresolved CO emission contribution falls within our extraction aperture and affects the extracted spectrum.

However, whether this effect can operate at the required level remains uncertain. The key question is whether the putative CO emission component can plausibly increase by a factor of $\gtrsim2$ during the burst---i.e., significantly faster than the continuum---such that emission filling could dominate the observed change in the normalized absorption depth. If the relevant emitting component persists across multiple burst cycles (e.g., a pre-existing wind/outflow contribution), then the incremental contribution from a single burst episode may be diluted by the baseline emission, making large-amplitude changes in the aperture-integrated CO emission less likely.

Additional constraints come from other emission tracers. The LTE analysis in Paper~III indicates that the characteristic temperature of the CO-emitting component is comparable to that of the H$_2$ emission detected with MIRI, suggesting that the two tracers may arise from physically related warm molecular gas in the outflow/cavity environment. Moreover, in our central-aperture spectra, the MIRI H$_2$ 0--0 S(3) (9.665\,\um) line flux is consistent within the uncertainties between the quiescent and burst epochs, and none of the other detected emission lines shows evidence for flux variations at the level of a factor of two. Taken together, these considerations disfavor a scenario in which burst-enhanced CO emission within the aperture increases by $\gtrsim2\times$ and dominates the apparent variability. We therefore regard emission filling as a plausible secondary contribution, but unlikely to be the dominant cause of the observed factor-of-two change in the normalized CO fundamental absorption depth.

\subsubsection{Scenario 4: Continuum emission change as the primary driver}
\label{subsubsec:scen_veil}

A fourth explanation is that the intrinsic absorption profile remains approximately stable while the continuum increases during the burst, reducing the normalized absorption depth through dilution (i.e., veiling). 
This interpretation is particularly attractive because the \water\ absorption bands show variability analogous to CO: both the CO fundamental and the \water\ bending-mode absorption become shallower during the burst. A continuum-driven dilution mechanism can therefore reduce the normalized depths of multiple molecular absorption bands simultaneously without requiring large intrinsic changes in the CO and \water\ column densities. This motivates the empirical veiling formalism adopted in Section~\ref{subsec:analysis_veiling ratio} and the physical interpretation of the wavelength-dependent behavior in terms of changes in the hot/warm continuum components discussed in Section~\ref{subsec:discussion_continuum}.

\subsubsection{Summary of constraints}
\label{subsubsec:scen_summary}

In summary, (1) a genuine decrease in the CO column density is disfavored by the isotopologue behavior and by the expected chemical resilience of CO in the inner gaseous disk \citep{Najita2007,Thi2005}; (2) changes in the effective $T_{\rm bg}/T_{\rm gas}$ ratio can provide an illustrative description of the depth change but still require order-unity dilution (e.g., $r_{4.6\,\mu m}\approx0.8$) and more detailed inner-disk modeling to be applied quantitatively; (3) emission filling within the extraction aperture is possible but unlikely to vary by $\gtrsim2\times$ and dominate the observed change, consistent with the lack of strong variability in aperture-extracted H$_2$ emission and with the outflow/cavity emission context established in Paper~III. Accordingly, we adopt continuum dilution as the primary interpretation for the observed CO fundamental variability, and we discuss its physical origin in the following section.

\subsection{Continuum Origin and Implications for Relative Veiling}
\label{subsec:discussion_continuum}

An essential ingredient in interpreting the absorption-line variability in EC\,53 is the physical origin of the infrared continuum that backlights the molecular gas. 
In protostars, the accretion luminosity can be released through both viscous dissipation in the inner disk and magnetospheric accretion shocks \citep{Bouvier2007}. 
However, in the infrared wavelength range relevant to the CO and \water\ rovibrational bands (2--8~\micron), our analysis indicates that the continuum emission is dominated by the viscously heated inner disk rather than by magnetospheric shocks.

We distinguish between ``hot'' and ``warm'' continuum components based on the characteristic temperatures of the absorbing gas traced by the molecular lines. The CO overtone band and the \water\ stretching mode trace gas at $T\sim1800$~K, while the CO fundamental band and the \water\ bending mode probe cooler gas at $T\sim1200$~K. These temperatures provide physically motivated lower limits on the temperature of the underlying continuum that backlights the absorption, rather than a strict decomposition of emission sources. 

In particular, although we compute the temperature profile $T(r)$ using only viscous heating (Equation~\ref{eq:T_vis}), the ``warm'' continuum component can be powered not only by viscous dissipation but also by passive heating due to irradiation from the hotter innermost regions (e.g., the hot inner disk, magnetospheric accretion shock, and the protostellar surface). A fully self-consistent treatment of these coupled heating sources and their impact on the emergent continuum requires detailed radiative-transfer modeling, which is beyond the scope of this work.

Within this framework, the enhanced relative veiling observed in the CO fundamental band during the burst phase can be understood in terms of the different heating mechanisms that contribute to the warm continuum. In the quiescent phase, the warm disk emission is expected to be dominated primarily by passive heating by the radiation originating from the hotter inner region. To zeroth order, if the disk’s reprocessing geometry (e.g., flaring/illumination and optical depth) does not change dramatically between epochs, the fraction of hot radiation intercepted and re-emitted by the warm disk should scale approximately with the hot component, yielding a broadly similar hot-to-warm continuum ratio. We note, however, that the reprocessing efficiency can be sensitive to the disk’s vertical structure and illumination, so a fully self-consistent prediction would require detailed radiative-transfer modeling.

During the burst phase, however, an additional heating channel becomes important. As the mass accretion rate through the disk increases, viscous dissipation in regions with midplane temperatures below $\sim1200$~K can increase substantially, directly boosting the intrinsic emission from the warm disk itself. This additional viscous heating increases the warm continuum contribution beyond what is expected from passive reprocessing alone, leading to a substantially larger warm-to-hot continuum ratio during the burst. As a result, absorption features in the CO fundamental band and the \water\ bending mode appear more strongly diluted in the burst phase, producing the observed increase in relative veiling.

Model-based estimates of the accretion energy budget and representative infrared continuum fluxes, including the relative roles of viscous disk emission and magnetospheric accretion shocks, are presented in Appendix~\ref{app:continuum}. These calculations support the qualitative interpretation adopted here and demonstrate that, even when magnetospheric accretion contributes substantially to the bolometric luminosity, the infrared continuum that backlights the molecular absorption lines is primarily dominated by disk emission.

\subsection{Spatial variation of Mass Accretion Rates}\label{subsec:dis_mdot}
\begin{figure}[!tp]
 \centering
 \vspace{-2mm}
 \includegraphics[width=0.45\textwidth]{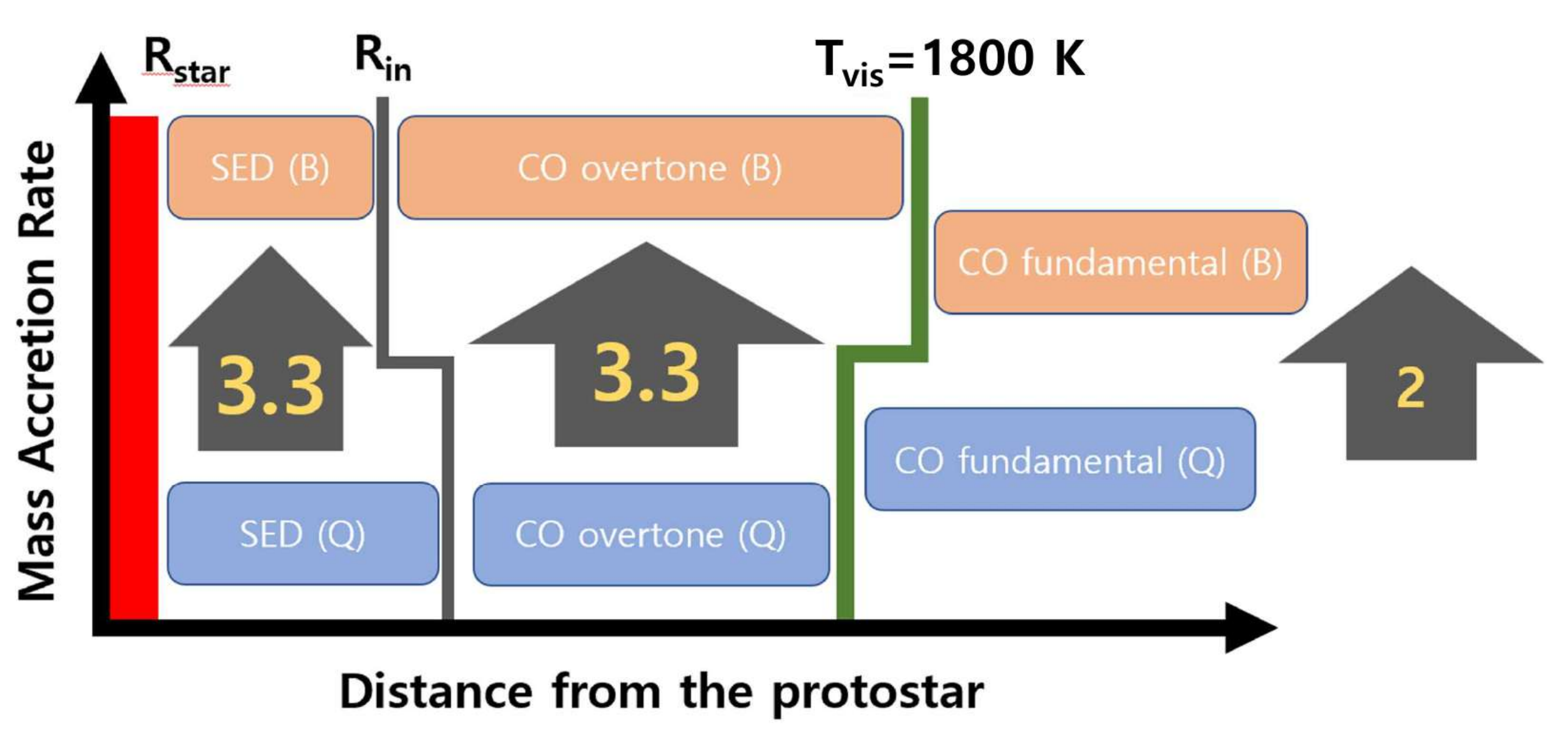}
 \vspace{-5mm}
 \caption{Schematic diagram of mass accretion rates in EC~53. The burst-to-quiescent mass-accretion-rate ratios are 3.3 based on SED modeling \citep{Baek2020} and CO overtone bands, and 2.0 based on CO fundamental lines. Assuming the total mass transport per cycle is conserved across all radii, the mass accretion rate through the outer disk must be higher than that of the inner disk during the quiescent phase, and lower during the burst phase,  as marked with orange and blue boxes. The vertical lines indicate the inner disk radius derived from Equation~(\ref{eq:r_in}) and the location corresponding to $T = 1800$\,K from Equation~(\ref{eq:T_vis}), both of which differ between the burst and quiescent phases.
}
\label{fig:mdot}
\end{figure}
In this section, we interpret the spatial variation of the accretion-rate enhancement between the burst and quiescent phases of EC\,53. Section~\ref{subsec:analysis_mdot} showed that the inferred burst-to-quiescent hot-midplane continuum ratio, and the corresponding $\dot{M}(\mathrm{B})/\dot{M}(\mathrm{Q})$, differ between bands tracing hotter ($T\gtrsim1800$\,K) and cooler ($T\gtrsim1200$\,K) components.  
While these ratios are derived from a simplified LTE slab model and depend on several assumptions, they are qualitatively consistent with a scenario in which the burst-related increase of the mass accretion rate varies with radius. We emphasize that this interpretation is tentative and should be regarded as a proof-of-concept rather than a definitive measurement of a radial dependence in $\dot{M}$. We therefore adopt a simple toy model to examine whether the inferred ratios are qualitatively consistent with such a scenario.

A possible physical origin for this behavior can be understood in the context of non-steady disk accretion. Previous studies of EC\,53 have suggested that episodic accretion arises from an imbalance between the mass inflow through the disk and the accretion onto the protostar, leading to mass buildup in the inner disk followed by rapid draining events \citep{yhLee2020}. In this picture, the accretion flow is intrinsically non-uniform with radius, as material can accumulate at specific disk locations before being released. Recent radiation-hydrodynamic simulations further provide a physical framework for such behavior by showing that material can pile up near the inner dead-zone edge and later be released through thermally triggered MRI activation \citep{Cecil2024}. More generally, theoretical studies have shown that the efficiency of angular momentum transport—and thus the accretion rate—can vary significantly with radius depending on local disk conditions such as ionization, temperature, and the activation of instabilities \citep[e.g.,][]{Armitage2011}, naturally leading to spatially varying accretion rates and episodic behavior.

In this context, the wavelength-dependent continuum variations presented in this work may provide a qualitative indication of such non-uniform accretion within the inner disk. We stress, however, that our analysis does not directly measure $\dot{M}(R)$, but rather tests whether the observed spectral variability is consistent with this physical picture.

As an empirical anchor for the innermost accretion-powered region, we adopt the burst and quiescent accretion rates inferred from bolometric luminosity/SED modeling. Following common usage, we refer to these as the accretion rate onto the protostar, $\dot M_*$, with a burst-to-quiescent ratio of 3.3: $\dot M_*(B)=6.7\times10^{-6}\,M_\odot\,{\rm yr^{-1}}$ and $\dot M_*(Q)=2.0\times10^{-6}\,M_\odot\,{\rm yr^{-1}}$ \citep{Baek2020,yhLee2020}\footnote{The mass accretion rate in \citet{yhLee2020} is $8.4\times10^{-6}\,M_\odot\,\mathrm{yr^{-1}}$ during the burst phase; adopting $R_{*}=1.6\,R_\odot$ instead of their assumed $2.0\,R_\odot$ reduces this value accordingly.}. We note, however, that the accretion-powered luminosity budget in EC\,53 is expected to be dominated by emission from the magnetospheric accretion shock together with the hottest inner disk (see Table~\ref{tb:luminoisity}); thus the SED-inferred $\dot M_*$ likely reflects the instantaneous accretion power of the innermost region and may include contributions associated with the hottest inner disk. Moreover, the CO overtone-based ratio, $\dot{M}_{\rm disk}(B)/\dot{M}_{\rm disk}(Q)=3.6\pm0.5$ (Table~\ref{tb:ratios}), is consistent within the uncertainties with the SED-based ratio. We therefore treat the SED-inferred $\dot M_*$ as representative of the accretion rate through the innermost hot disk traced by the CO overtone ($T\gtrsim1800$\,K) for the purposes of the toy analysis below. By contrast, the CO fundamental band implies a lower burst-to-quiescent ratio of $\sim2.0$ at somewhat larger radii within the inner disk. In the discussion below, $\dot M_{\rm disk}$ should therefore be interpreted as a representative (band-weighted) accretion rate for the disk region traced by a given molecular band, rather than a unique local value at a single radius.

Using the fiducial inner-disk ratio together with the lower ratio of $\sim2.0$ inferred from the CO fundamental band, we estimate characteristic accretion rates at larger radii in a time-averaged sense over the burst cycle. Because EC\,53 exhibits periodic luminosity bursts \citep{yhLee2020}, we assume that, over one $\sim$1.5\,yr cycle, the cycle-averaged mass transport through the disk balances the mass accreted onto the star. Adopting the luminosity curve of \citet[][their Eq.~23]{yhLee2020} together with a burst-to-quiescent ratio of 3.3, we approximate the cycle as a step function in which the burst occupies 36~\% and the quiescent phase 64~\% of the period. Applying the inferred larger-radius ratio of 2.0 then yields characteristic accretion rates of $5.4\times10^{-6}\,M_\odot\,{\rm yr^{-1}}$ (burst) and $2.7\times10^{-6}\,M_\odot\,{\rm yr^{-1}}$ (quiescence) for the cooler inner-disk region traced by the CO fundamental band.

Figure~\ref{fig:mdot} summarizes this toy interpretation of the inferred accretion-rate ratios as a function of radius during the burst and quiescent phases. In this picture, the quiescent-phase $\dot M_{\rm disk}$ at larger radii exceeds the fiducial inner rate anchored by the SED, whereas during the burst the larger-radius $\dot M_{\rm disk}$ falls below it. This contrast suggests that material can accumulate in the inner disk during quiescence and be more efficiently drained through the innermost region during the burst, qualitatively consistent with episodic accumulation and release in the inner disk around $\sim$0.05\,au proposed by \citet{yhLee2020}.

We caution that this toy interpretation is highly simplified, and the inferred accretion-rate ratios should be regarded as band-weighted, phenomenological constraints rather than a unique reconstruction of $\dot M(r)$. 
First, our translation from the hot-continuum ratios to $\dot M(B)/\dot M(Q)$ is anchored to temperature thresholds (e.g., $T>1800$\,K and $T>1200$\,K), but the radii at which a given temperature is reached shift between epochs (e.g., the $T=1800$\,K radius moves from $\sim$0.06\,au in quiescence to $\sim$0.09\,au in burst). As a result, the ratios inferred from different bands do not map one-to-one onto fixed spatial locations, complicating any direct interpretation in terms of a unique radial profile $\dot M(r)$.
Second, while we use the CO overtone and fundamental bands to represent ``inner'' and ``somewhat larger-radius'' tracers, the CO fundamental absorption is not confined to a narrow annulus exterior to the overtone region; instead, it is contributed by all gas hotter than $\sim$1200\,K and therefore includes the hotter inner radii that also contribute to the overtone absorption. 
Third, although we discuss radius-dependent accretion behavior, the viscous-heating prescription used to relate $\dot M$ to $T(r)$ (Equation~\ref{eq:T_vis}) is derived for a steady disk with $\dot M={\rm const}$; relaxing this assumption and treating a radially varying $\dot M$ self-consistently requires a full radiative-transfer calculation coupled to a time-dependent disk-evolution model. 
Nevertheless, Figure~\ref{fig:mdot} provides a useful qualitative illustration of how the inferred epoch-to-epoch accretion-rate ratios may be interpreted as evidence for non-uniform accretion enhancement across the inner disk of EC\,53.

\subsection{Accretion Variability in EC~53 and Comparison with Other Episodic Accretors}

Previous monitoring studies have already shown that EC~53 is unusual among episodic accretors in combining a deeply embedded Class~I geometry with remarkably regular burst cycles. Its luminosity bursts repeat on a timescale of $\sim$530 days, with brightenings of $\sim$1.5 mag in the $K$ band, $\sim$2 mag at 3.35\,\um, and only $\sim$0.3 mag at 850\,\um\ \citep{Hodapp2012,Yoo2017,yhLee2020}. This strong wavelength dependence is expected if the shorter-wavelength emission responds more directly to changes in the central accretion-powered luminosity, whereas the submillimeter flux primarily traces the weaker temperature response of the surrounding envelope. Consistent with this picture, radiative-transfer modeling showed that the observed factor of $\sim$1.5 brightening at 850\,\um\ corresponds to an internal luminosity increase by a factor of $\sim$3.3 rather than a one-to-one change in the observed continuum flux \citep{Baek2020}. In the picture proposed by \citet{yhLee2020}, the quasi-periodic variability arises from cyclic filling and draining of the inner disk reservoir, while companion-triggered or companion-modulated accretion remains a plausible additional interpretation rather than a uniquely established mechanism.

Our results add a spectroscopic view of this accretion variability. The burst-to-quiescent ratio inferred from the CO overtone band, $\dot{M}(B)/\dot{M}(Q)=3.6\pm0.5$, is broadly consistent with the internal-luminosity increase inferred from the bolometric luminosity/SED analysis and therefore appears to trace the accretion-powered continuum of the innermost hot disk region. By contrast, the CO fundamental band implies a lower representative ratio of $\sim$2.0 for a somewhat cooler continuum-emitting region. As discussed in Section~\ref{subsec:dis_mdot}, however, these values should be interpreted as band-weighted, phenomenological estimates rather than as a unique reconstruction of a radial accretion-rate profile.

This distinction is important when comparing the inferred ratios with the observed photometric variability. The representative accretion-rate ratios derived from the CO overtone and fundamental bands should not be interpreted as a single global accretion-rate ratio for the whole system, nor compared directly with the observed factor of $\sim$4 photometric variability. Instead, they characterize the accretion-related continuum variations in the specific backlighting regions traced at those wavelengths. In particular, the continuum at each wavelength may include not only locally generated hot emission but also reprocessed radiation from shorter wavelengths. As a result, the observed continuum ratios and the inferred accretion-rate ratios need not scale in a one-to-one way. This is illustrated in Table~\ref{tb:ratios}: for the CO overtone band, negligible relative veiling allows the observed continuum ratio to track the hot-continuum ratio directly, whereas for the CO fundamental band the larger relative veiling implies that the observed continuum ratio of $\sim$3.3 corresponds to a smaller hot-continuum ratio of $\sim$1.7 and hence to a lower representative accretion-rate ratio of $\sim$2.0.

Placed in the broader landscape of episodic accretors, EC~53 appears to occupy an intermediate regime in both absolute accretion level and burst amplitude. EX~Lup provides a lower-accretion, large-amplitude benchmark in an emission-dominated Class~II disk, with strong molecular and chemical variability during outburst \citep{Goto2011,Banzatti2012,Banzatti2015,Smith2025}. DQ~Tau offers a useful comparison for periodic accretion modulation: its accretion rate varies between $\sim10^{-8.5}$ and $\sim10^{-7.3}\,M_\odot\,{\rm yr}^{-1}$ over the orbit \citep{Fiorellino2022}, and recent JWST monitoring shows that some hot molecular components correlate strongly with accretion state while warmer or colder components can remain comparatively stable \citep{Kospal2025}. At the opposite extreme, FU~Ori-type systems reach burst accretion rates of order $\sim10^{-5}$ to a few $\times10^{-4}\,M_\odot\,{\rm yr}^{-1}$ and commonly exhibit continuum-dominated spectra in absorption \citep{Audard2014}. EC~53 therefore differs from both the EXor-like and FUor-like extremes: it shows only moderate representative burst-to-quiescent accretion-rate changes, but at a relatively high absolute accretion level in a deeply embedded Class~I environment. In this sense, EC~53 occupies an intermediate regime, linking lower-accretion periodic systems such as DQ~Tau with the more extreme eruptive behavior seen in EX~Lup and FUor-type objects.

\section{Summary}\label{sec:summary}
We present JWST NIRSpec and MIRI observations of EC~53 obtained during both the quiescent and burst phases, covering the CO overtone and fundamental bands as well as the \water\ stretching and bending modes. All four sets of rovibrational lines are detected in absorption. The current JWST CO overtone data do not provide a robust constraint on overtone variability, while the \water\ stretching-mode absorption shows no compelling evidence for variability between the two epochs. In contrast, the CO fundamental and \water\ bending-mode lines exhibit significant reductions in normalized depth during the burst phase. This pattern is most naturally explained by changes in the continuum emission between the two phases, rather than by large variations in the absorbing gas.

A simple LTE slab analysis suggests that the CO overtone and fundamental bands trace gas at temperatures of $\sim$1800\,K and $\sim$1200\,K, respectively, implying that the overtone absorption probes hotter gas at smaller  radii than the fundamental absorption. To quantify the epoch-to-epoch change in continuum dilution, we introduce the relative veiling, $\mathcal{V}$, which uses the quiescent spectrum as an internal reference and measures continuum dilution directly from the transformation between the two normalized spectra.  This internal formalism enables us to link the observed $\mathcal{V}$ to the underlying hot-continuum emission that backlights the absorption. Within a simplified framework, we interpret the inferred hot-continuum variation as arising from a viscously heated, optically thick inner disk, and use it as a phenomenological diagnostic of accretion-powered variability. Using this framework, we infer representative burst-to-quiescent hot-continuum ratios of $2.9\pm0.2$ for the inner-disk region traced by the CO overtone band and $1.71\pm0.11$ for the somewhat cooler region traced by the CO fundamental band.
These values provide the most direct observational measure in this work of how the accretion-powered inner-disk continuum changes between the two phases.

We then translate these hot-continuum ratios into representative burst-to-quiescent mass-accretion-rate ratios using a simplified viscous-heating prescription for the inner disk.
We find that the burst-to-quiescent accretion-rate ratio is larger for the hottest inner-disk component (as traced by the CO overtone band; $3.6\pm0.5$) than for the cooler component at somewhat larger radii traced by the CO fundamental band ($\sim2.0$). Within the uncertainties, the inner-disk ratio is consistent with the ratio of $\sim$3.3 inferred from the bolometric luminosity/SED \citep{Baek2020}. These inferred accretion-rate ratios are model-dependent and should be regarded as representative rather than uniquely determined values.

Interpreting these ratios in a simple periodic-cycle picture, where the time-averaged mass transport over a full cycle must be balanced, implies a reversal in the radial ordering of the accretion rate between phases (Figure~\ref{fig:mdot}): during quiescence the accretion rate at larger radii exceeds that through the innermost region, whereas during the burst the innermost accretion rate is enhanced relative to larger radii. This behavior supports the episodic accumulation-and-release scenario proposed by \citet{yhLee2020}. We note, however, that this is a proof-of-concept interpretation rather than a direct measurement of a unique radial accretion-rate profile. Additional multi-epoch monitoring and complementary high-resolution spectroscopy, together with radiative-transfer calculations and detailed disk-evolution modeling, will be essential for investigating the accretion physics in EC~53.

\vspace{5mm}
This work is supported by the Korea Astronomy and Space Science Institute
under the R\&D program (project No. 2026-1-833-01) supervised by the Ministry of Science and ICT.  This work was supported by the National Research Foundation of Korea (NRF) grant funded by the Korea government (MSIT) (grant numbers RS-2024-00416859 and RS-2026-25490557). G.J.H. is supported by grant IS23020 from the Beijing Natural Science Foundation. D.J. is supported by NRC Canada and by an NSERC Discovery Grant. G.B. was supported by Basic Science Research Program through the National Research Foundation of Korea (NRF) funded by the Ministry of Education (RS-2023-00247790).  J.D.G. acknowledges support from the associated  3477 NASA observer grant. 
This work is based on observations made with the NASA/ESA/CSA James Webb Space Telescope. The data were obtained from the Mikulski Archive for Space Telescopes at the Space Telescope Science Institute, which is operated by the Association of Universities for Research in Astronomy, Inc., under NASA contract NAS 5-03127 for JWST. These observations are associated with the JWST program \#3477. 
This work used the Immersion Grating Infrared Spectrograph 2 (IGRINS-2) developed and built by a collaboration between KASI and the International Gemini Observatory. Based on observations obtained at the international Gemini Observatory, a program of NSF NOIRLab, which is managed by the Association of Universities for Research in Astronomy (AURA) under a cooperative agreement with the U.S. National Science Foundation on behalf of the Gemini Observatory partnership: the U.S. National Science Foundation (United States), National Research Council (Canada), Agencia Nacional de Investigación y Desarrollo (Chile), Ministerio de Ciencia, Tecnología e Innovación (Argentina), Ministério da Ciência, Tecnologia, Inovações e Comunicações (Brazil), and Korea Astronomy and Space Science Institute (Republic of Korea). This work was enabled by observations made from the Gemini North telescope, located within the Maunakea Science Reserve and adjacent to the summit of Maunakea. We are grateful for the privilege of observing the Universe from a place that is unique in both its astronomical quality and its cultural significance. 
Nayuta, Kagoshima, and IRSF telescopes are partially supported
by the Optical and Infrared Synergetic Telescopes for Education
and Research (OISTER) program funded by the MEXT of Japan. The JWST data presented in this article were obtained from the Mikulski Archive for Space Telescopes (MAST) at the Space Telescope Science Institute. The specific observations analyzed can be accessed via \dataset[doi:10.17909/h56s-xr27]{https://doi.org/10.17909/h56s-xr27}.
We acknowledge using ChatGPT to check English grammar and improve the clarity of expressions. 

\appendix
\restartappendixnumbering

\section{NIR Photometry}
\label{sec:nir_photometry}

\begin{figure*}[htp]
 \centering
 \vspace{-2mm}
 \includegraphics[width=0.9\textwidth]{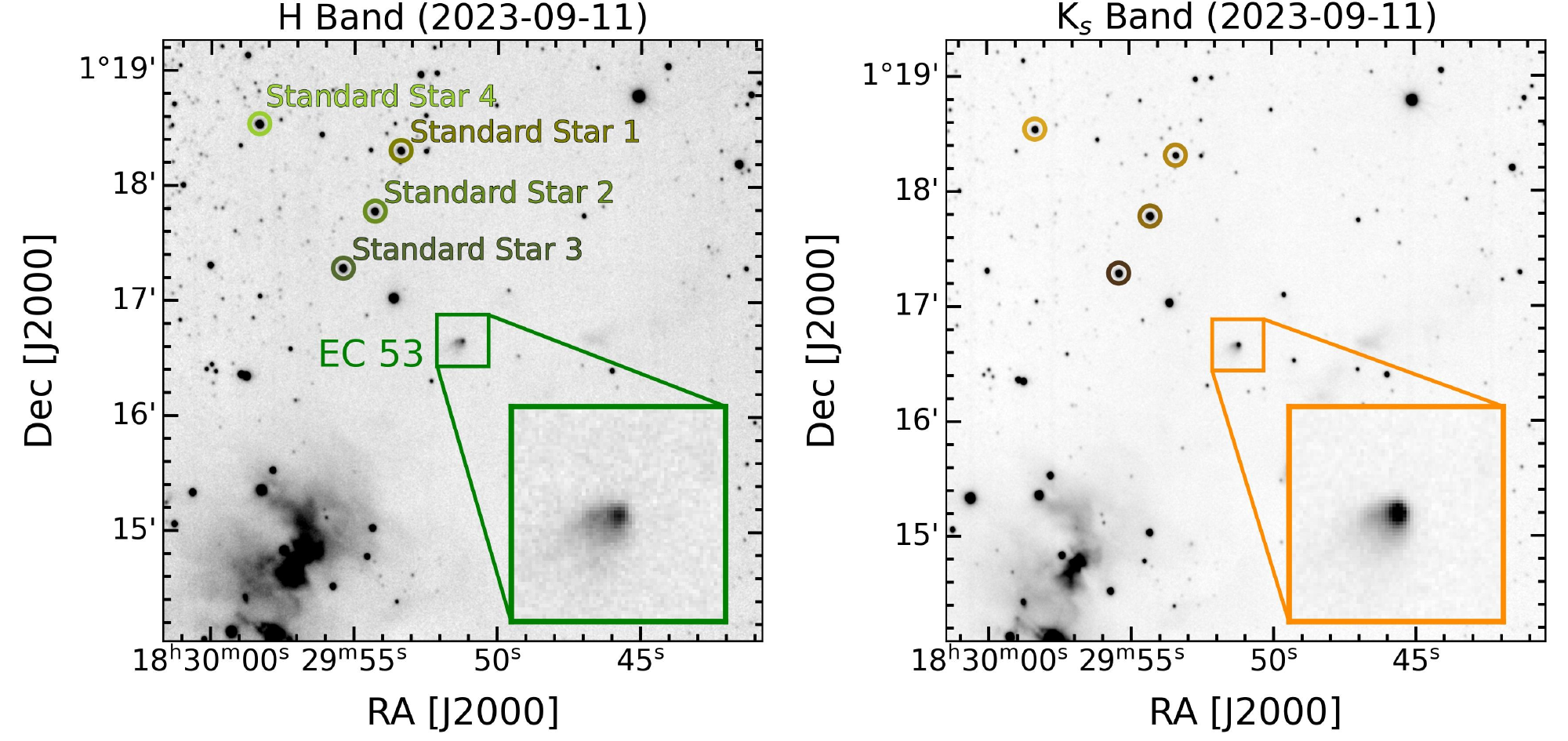}
 \caption{Photometric images of EC 53 obtained with IRSF in the H band (\textit{left}) and Ks band (\textit{right}). The position of EC 53 is marked by green and orange boxes in the H- and $K_s$ bands, respectively. Zoomed-in views of EC~ 53 are shown in the lower right corner of each panel. The positions of the standard stars used for photometric calibration are denoted by olive and brown circles. }
\label{fig:irsf_phot}
\end{figure*}

\begin{deluxetable*}{cccccc}
\tablecaption{Standard Stars for NIR Photometry \label{tb:standstar}}
\tablehead{
\colhead{\bf Name$^a$} & \colhead{\bf Target} & \multicolumn{2}{c}{\bf Position [J2000]} &\multicolumn{2}{c}{\bf Magnitude$^b$  [mag]}  \\
\colhead{} & \colhead{} & \colhead{R.A.}& \colhead{Dec.}& \colhead{H}& \colhead{Ks} }
\startdata
Standard Star 1& J18295338+0118185  & 18h 29m 53.38s & +1\degree 18$^\prime$ 18.54\arcsec & 12.478 & 11.512 \\
Standard Star 2& J18295428+0117468 & 18h 29m 54.28s & +1\degree 17$^\prime$ 46.82\arcsec & 12.409 & 10.478  \\
Standard Star 3& J18295538+0117170 & 18h 29m 55.39s & +1\degree 17$^\prime$ 17.07\arcsec & 12.147  & 10.654  \\
Standard Star 4& J18295831+0118322 & 18h 29m 58.32s & +1\degree 18$^\prime$ 32.27\arcsec & 11.867 & 11.155 \\
 \enddata
\tablecomments{$^a$ Identification shown in Figure~\ref{fig:irsf_phot}\\
            $^b$ Magnitudes adopted from 2MASS catalog} 
\end{deluxetable*}

The NIR light curve of EC 53 (Figure~\ref{fig:lightcurve}) was constructed from monitoring observations obtained with the Nayuta, Kagoshima and IRSF telescopes. Figure~\ref{fig:irsf_phot} shows representative H and $K_s$-band images of EC~53 obtained with the IRSF telescope during a quiescent phase. 
Aperture photometry was performed using the Source Extractor \citep{Bertin1996}. The aperture size for each source was automatically determined with Kron's first moment algorithm \citep{Kron1980}, and fluxes were measured only for detections with SNRs greater than 3. 

Instrumental magnitudes were calibrated via differential photometry by cross-matching detected sources with the 2MASS catalog. Four standard stars were selected for zero-point calibration based on the following criteria: (1) location within the common field of view of all three telescopes, (2) detection in both the H and $K_s$ bands, and (3) no evidence for variability in the WISE W1 and W2 bands. The adopted standard stars are listed in Table~\ref{tb:standstar}.

For the Nayuta telescope, multiple observations were carried out per day and combined images using images with overlapping field of view. Because the observed fields varied between observations, standard stars 1 and 4 did not always fall within the overlapping regions and were thus excluded from the calibration for these data. In this case, only standard stars 2 and 3 were used to determine the photometric zero point.

\begin{figure*}[htp]
 \centering
 \vspace{-2mm}
 \includegraphics[width=1.0\textwidth]{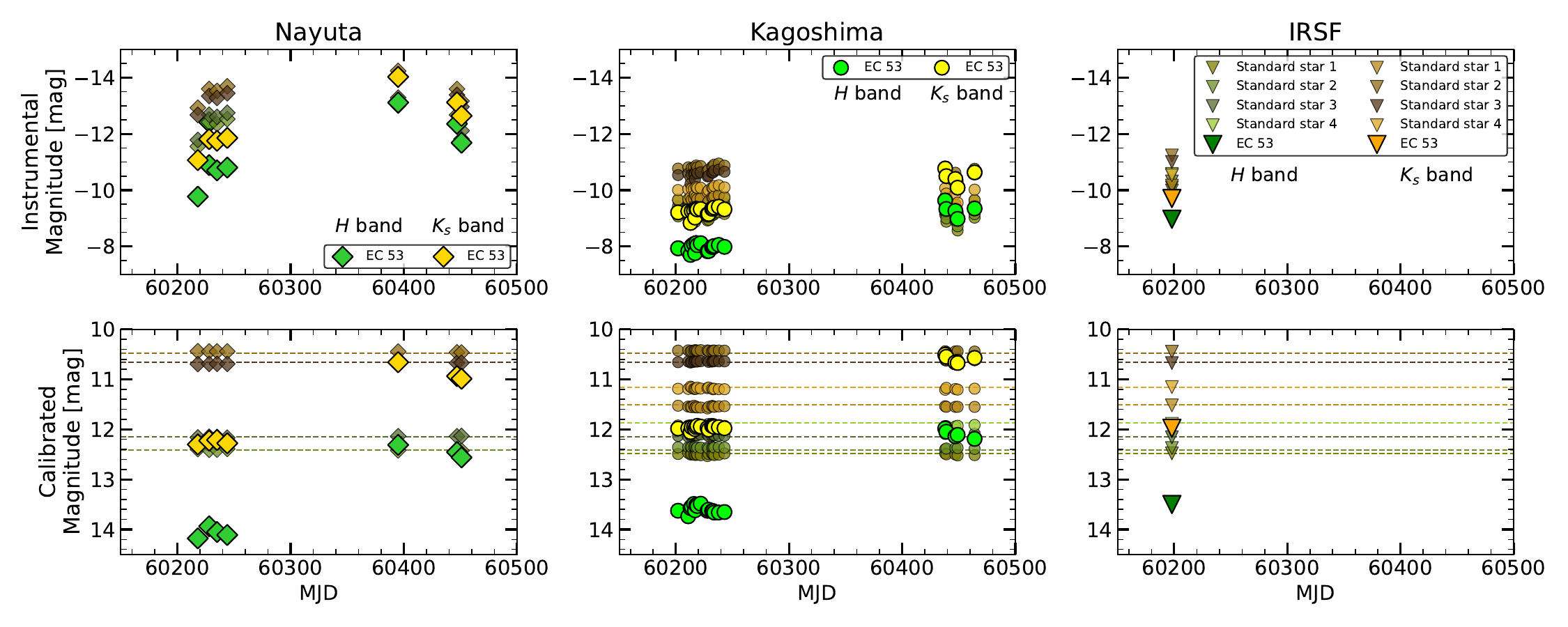}
 \caption{Magnitude calibration for the NIR monitoring observations obtained with the Nayuta (\textit{left}), Kagoshima University (\textit{middle}), and IRSF (\textit{right}) telescopes. The top panels show the instrumental magnitudes, while the bottom panels show the calibrated magnitudes derived from differential photometry. Green and yellow points represent the H- and $K_s$-band magnitudes of EC 53, respectively. Olive and brown points indicate the standard stars used for calibration. The dashed lines mark the corresponding magnitudes from the 2MASS catalog.}
\label{fig:phot_mag_cal}
\end{figure*}

\begin{figure*}[htp]
 \centering
 \vspace{-2mm}
 \includegraphics[width=1.0\textwidth]{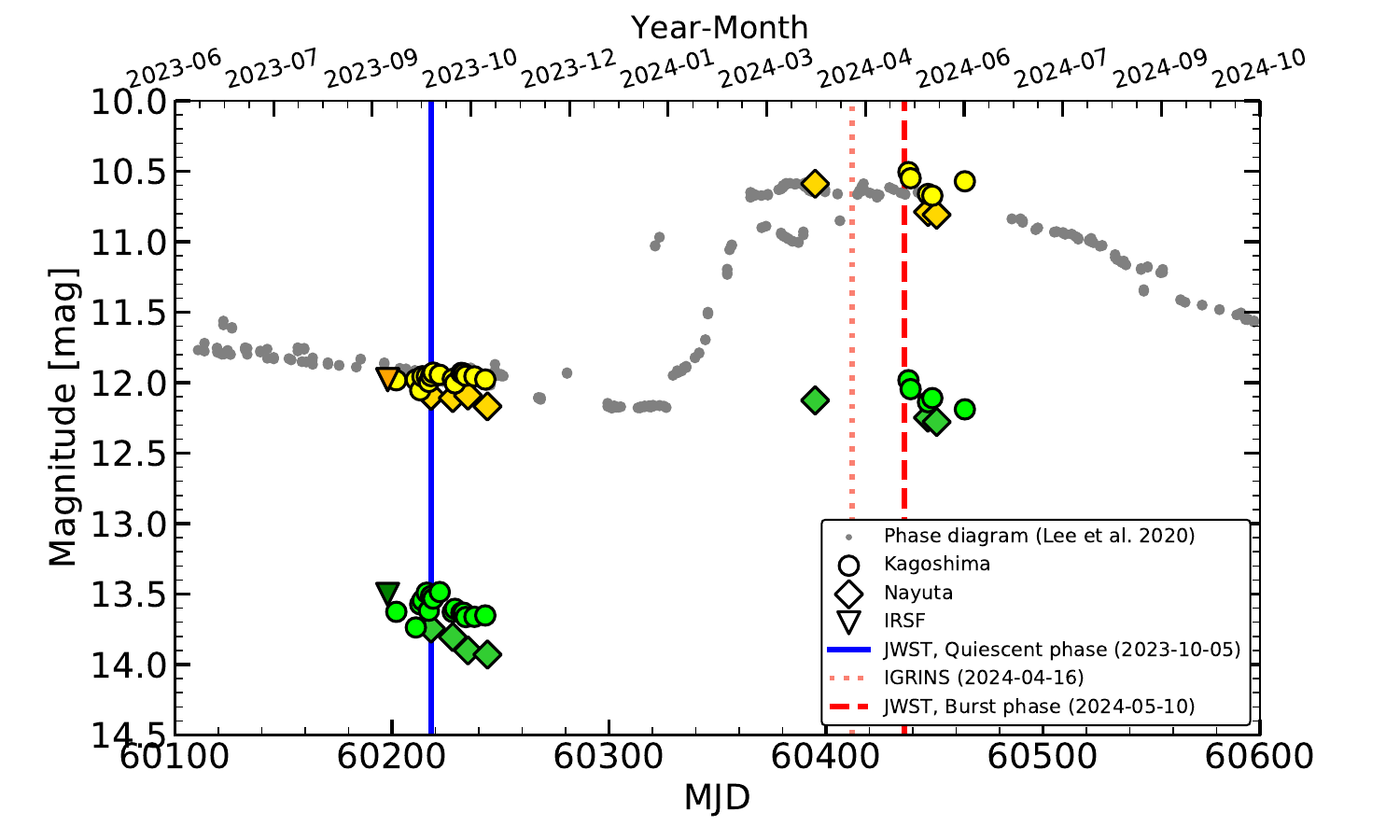}
 \caption{Same as  Figure \ref{fig:lightcurve}, but showing the H-band light curve, plotted with green points. }
\label{fig:lightcurve_hband}
\end{figure*}

Figure~\ref{fig:phot_mag_cal} shows the resulting magnitude calibration for the NIR light curves. The zero-point magnitude was determined as the average offset between the instrumental and 2MASS magnitudes of the selected standard stars.  As shown in Figure~\ref{fig:lightcurve_hband}, the magnitudes derived from the Nayuta data exhibit a small systematic offset relative to those from the other telescopes, likely reflecting the limited number of standard stars available for calibration in this case.

\section{LTE Slab Model for Absorption}
\label{app:lte_model}

We model the molecular absorption using a simple LTE slab model characterized by a single gas temperature and column density.

Under the assumption of local thermodynamic equilibrium (LTE), the population of each energy level is given by the Boltzmann distribution:

\begin{equation}
N_j = N_{\rm tot} \frac{g_j}{Q(T)} \exp\left(-\frac{E_j}{kT}\right),
\end{equation}

where $N_j$ is the column density of molecules in energy level $j$, $N_{\rm tot}$ is the total column density of the molecule, $Q(T)$ is the partition function at temperature $T$, $g_j$ is the statistical weight of level $j$, $E_j$ is the energy of level $j$, $k$ is the Boltzmann constant, and $T$ is the gas temperature.

The optical depth for each individual transition $i$ is computed as:

\begin{equation}
\tau_{\nu,i} = \frac{c^2}{8\pi \nu_i^2} A_{ul,i} \, N_l \left(1 - \frac{g_l N_u}{g_u N_l}\right) \phi_i(\nu),
\end{equation}

where $c$ is the speed of light, $\nu_i$ is the line center frequency of transition $i$, $A_{ul,i}$ is the Einstein A coefficient, $N_l$ and $N_u$ are the column densities of the lower and upper levels, and $g_l$ and $g_u$ are their statistical weights. The term in parentheses accounts for stimulated emission.

The line profile function $\phi_i(\nu)$ is assumed to be Gaussian:

\begin{equation}
\phi_i(\nu) = \frac{1}{\sqrt{\pi}\Delta \nu_D} 
\exp\left[-\left(\frac{\nu - \nu_i}{\Delta \nu_D}\right)^2\right],
\end{equation}

where $\Delta \nu_D$ is the Doppler line width corresponding to the adopted intrinsic velocity width.

The total optical depth is obtained by summing over all transitions:

\begin{equation}
\tau_\nu = \sum_i \tau_{\nu,i},
\end{equation}

thereby including line overlap among transitions of the same molecule. 

The emergent intensity is given by the radiative transfer equation:

\begin{equation}
I_\nu = I_{\nu,{\rm bg}} \exp(-\tau_\nu) + B_\nu(T) \left[1 - \exp(-\tau_\nu)\right],
\end{equation}

where $I_{\nu,{\rm bg}}$ is the background continuum intensity and $B_\nu(T)$ is the Planck function at the gas temperature $T$.

In the case considered here, the background continuum originates from the hot inner disk and has a temperature significantly higher than that of the absorbing gas. Under this condition, $I_{\nu,{\rm bg}} \gg B_\nu(T)$, and the second term becomes negligible. The emergent spectrum can therefore be approximated as:

\begin{equation}
I_\nu \simeq I_{\nu,{\rm bg}} \exp(-\tau_\nu).
\end{equation}

Since the observed spectra are normalized by the continuum level, the resulting normalized flux is given by:

\begin{equation}
\frac{I_\nu}{I_{\nu,{\rm bg}}} \simeq \exp(-\tau_\nu).
\end{equation}

Finally, the modeled spectra are convolved with the instrumental spectral resolution of each dataset before comparison with the observations.

\begin{figure}[!tp]
 \centering
 \vspace{-2mm}
 \includegraphics[width=0.45\textwidth]{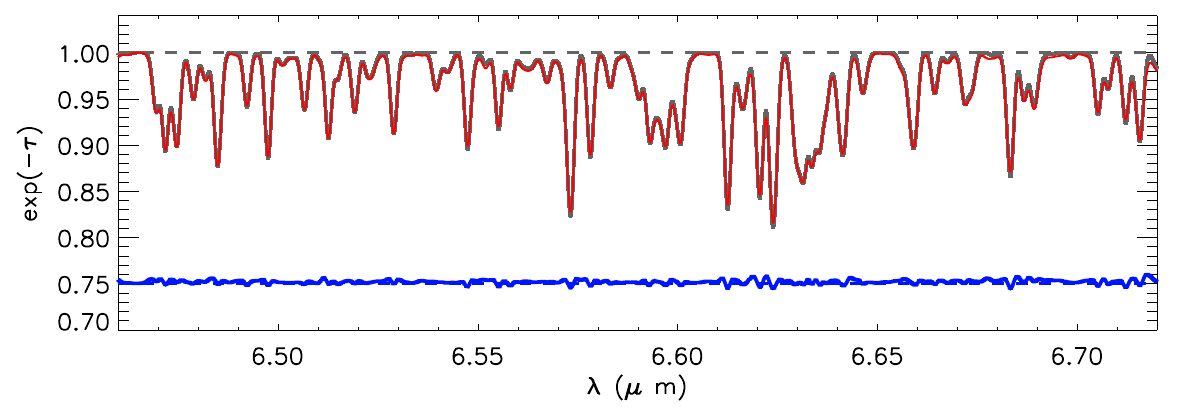}
 \caption{Comparison between our H$_2$O slab-model spectrum (red) and the publicly available model of \citet{Li2024} (black), computed with the same input parameters (see text). The blue curve, shown with a vertical offset, represents the residual difference between the two spectra. The overall agreement is very good, with only minor discrepancies.
}
\label{fig:comp_li_model}
\end{figure}

As a consistency check, we compared our slab-model spectrum with the publicly available H$_2$O absorption model of \citet{Li2024} using their online spectrum generator\footnote{\url{https://mirasg.astro.umd.edu/}} and the same input parameters: spectral resolution $R=3500$, excitation temperature $T_{\rm ex}=1200$\,K, column density $N=1\times10^{18}$\,cm$^{-2}$, fractional coverage $f_c=1.0$, line width ${\rm FWHM}=25$\,\kms, and velocity shift $\Delta v=0$\,\kms. As shown in Figure~\ref{fig:comp_li_model}, the two spectra are in very good agreement in both line positions and overall absorption depths, with only minor residual differences attributable to details of the numerical implementation. This comparison confirms that the two implementations produce equivalent results.

\section{Continuum Emission}
\label{app:continuum}

\begin{deluxetable}{crr}
\tablecaption{Modeled accretion energy budget and representative infrared continuum fluxes}
\label{tb:luminoisity}
\tablehead{
    \colhead{\bf Components} & \colhead{\bf Quiescence} & \colhead{\bf Burst}}
\startdata
\multicolumn{3}{l}{\it Energy budget ($L_\odot$)}\\
\hline
Bolometric luminosity$^{a}$     & 6.0 (100) & 20.0 (100)\\
Magnetospheric accretion shocks & 3.5 (58)  & 0.5 (2.5)\\
Disk ($T>1800$\,K)              & 1.7 (28)  & 10.4 (52)\\
Disk (1200\,K$<T<$1800\,K)      & 0.4 (6)   & 0.7 (3.5)\\
Disk ($T<1200$\,K)$^{b}$        & 0.3 (5)   & 8.4 (42)\\
\hline
\multicolumn{3}{l}{\it CO overtone continuum flux at 2.3\,\um\ (mJy)}\\
\hline
Magnetospheric accretion shocks & 28  & 4\\
Disk ($T>1800$\,K)              & 80  & 289\\
\hline
\multicolumn{3}{l}{\it CO fundamental continuum flux at 4.6\,\um\ (mJy)}\\
\hline
Magnetospheric accretion shocks & 9   & 1.3\\
Disk ($T>1200$\,K)              & 82  & 235\\
\enddata
\tablecomments{
$^{a}$ Total accretion luminosity adopted from \citet{Baek2020}.\\
$^{b}$ Residual luminosity after subtracting the listed hot components.\\
Percentages in parentheses indicate fractional contributions to the total accretion-powered energy budget.}
\end{deluxetable}

In protostars, the accretion luminosity is released primarily through viscous dissipation in the inner disk and, to a lesser extent, through magnetospheric accretion shocks \citep{Bouvier2007}. 
Both processes can, in principle, contribute to the continuum emission that backlights molecular absorption lines.
To place the interpretation adopted in the main text on a quantitative footing, we evaluate a model-based accretion energy budget and representative infrared continuum fluxes using the prescriptions described in Section~\ref{subsec:method_hot}.
The resulting estimates are summarized in Table~\ref{tb:luminoisity} and are intended to illustrate the relative importance of different accretion-powered components rather than to provide a detailed radiative-transfer solution.

For the purposes of this appendix, we separate the disk emission into ``hot'' and ``warm'' components using the characteristic temperatures traced by the absorption lines: $T\sim1800$\,K for the CO overtone band (2.3\,\um) and $T\sim1200$\,K for the CO fundamental band (4.6\,\um).
These temperatures provide physically motivated lower limits on the temperature of the midplane continuum that backlights each absorption band.
We emphasize that the warm component is not determined by viscous heating alone.
Passive reprocessing of radiation from the hotter inner disk, as well as stellar and shock emission, is expected to play an important role, particularly at lower temperatures.
A fully self-consistent treatment of this reprocessing requires detailed radiative-transfer modeling and is beyond the scope of this work.

\subsection{Hot continuum emission is disk-dominated in the infrared}

Although magnetospheric accretion shocks can contribute substantially to the bolometric accretion luminosity, their emission is concentrated primarily at ultraviolet and optical wavelengths.
In the infrared range relevant to this study (2.2--7.6\,\um), the continuum emission is instead dominated by the viscously heated inner disk.

In the adopted model, the luminosity of the magnetospheric accretion shock depends sensitively on the inner disk truncation radius (Equation~\ref{eq:r_in}).
As the mass accretion rate increases, the truncation radius approaches the stellar surface ($R_{\rm in}\approx1.4\,R_*$ in quiescence and $\approx1.0\,R_*$ in the burst phase), causing the shock luminosity to peak at $\dot{M}\sim3\times10^{-6}\,M_\odot\,\mathrm{yr}^{-1}$ and then decline at higher accretion rates.
As a result, the shock contribution decreases from $\sim3.5\,L_\odot$ in quiescence to $\sim0.5\,L_\odot$ during the burst phase, despite the higher overall mass accretion rate.

Crucially, this behavior does not alter the dominance of the disk in the infrared.
For example, at 2.3\,\um\ in quiescence, the magnetospheric shock contributes $\sim28$\,mJy, whereas the hot disk component ($T>1800$\,K) contributes $\sim80$\,mJy.
Even in conservative cases where the total shock luminosity is comparable to or exceeds that of the hot disk, the infrared continuum remains dominated by disk emission by factors of $\gtrsim3$.
This confirms that the hot midplane continuum backlighting the molecular absorption lines is primarily set by viscous heating of the inner disk rather than by magnetospheric accretion shocks.

\subsection{Warm continuum enhancement during the burst phase}

Within this framework, the accretion-powered energy budget provides a natural explanation for the enhanced role of warm continuum emission during the burst phase, particularly at wavelengths relevant to the CO fundamental band.
As the total accretion luminosity increases, a larger fraction of the accretion power is dissipated at radii where the disk temperature falls below $\sim1200$\,K, increasing the energetic importance of warmer disk regions.

As summarized in Table~\ref{tb:luminoisity}, the innermost hot region dominates the accretion-powered energy output in quiescence, whereas during the burst a substantially larger fraction of the energy budget is associated with lower-temperature regions of the disk.
We stress that these values describe the distribution of accretion energy rather than directly observable infrared fluxes.
Passive reprocessing by the dusty disk is expected to redistribute a significant fraction of the energy released at small radii toward longer wavelengths.

An enhanced warm continuum during the burst phase naturally dilutes molecular absorption features, leading to a larger apparent reduction in normalized line depths—and hence a larger relative veiling—in the CO fundamental band and the \water\ bending mode.
In contrast, the CO overtone band and the \water\ stretching mode remain dominated by emission from the hottest ($T\gtrsim1800$\,K) inner disk regions, where the contribution from warmer material is comparatively minor.




\end{document}